\documentstyle[12pt,titlepage,epsfig]{article}
\def\slash#1{\setbox0=\hbox{$#1$}#1\hskip-\wd0\dimen0=5pt\advance
       \dimen0 by-\ht0\advance\dimen0 by\dp0\lower0.5\dimen0\hbox
         to\wd0{\hss\sl/\/\hss}}
\def\bbuildrel#1_#2^#3{\mathrel{\mathop{\kern 0pt#1}\limits_{#2}^{#3}}}
\newcommand{\scs}{\scriptscriptstyle}
\newcommand{\be}{\begin{equation}}
\newcommand{\ee}{\end{equation}}
\newcommand{\bea}{\begin{eqnarray}}
\newcommand{\eea}{\end{eqnarray}}
\newcommand{\f}{\frac}

\newcommand{\al}{\alpha_s}

\newcommand{\Bsg}{$B{\to}X_s\gamma $ }

\begin{document}
\begin{titlepage}

 \begin{flushright}
  {\bf TUM-HEP-321/98\\       
       IFT-15/98\\
       CERN-TH/99-88\\
      hep-ph/9904413\\
}
\end{flushright}

 \begin{center}
  \vspace{0.6in}

\setlength {\baselineskip}{0.3in}
  {\bf \Large 
Matching Conditions for $b \to s \gamma$ and $b \to s \; gluon$\\
in Extensions of the Standard Model}
\vspace{2cm} \\
\setlength {\baselineskip}{0.2in}

{\large  Christoph Bobeth$^{^{1}}$, 
         Miko{\l}aj Misiak$^{^{2,3,\star}}$
         and J{\"o}rg Urban$^{^{1}}$}\\

\vspace{0.2in}
$^{^{1}}${\it Institut f{\"u}r Theoretische Physik, Technische Universit{\"a}t Dresden, \\
                        Mommsenstr. 13, D-01062 Dresden, Germany}

\vspace{0.2in}
$^{^{2}}${\it Physik Department, Technische Universit{\"a}t M{\"u}nchen,\\
                         D-85748 Garching, Germany}

\vspace{0.2in}
$^{^{3}}${\it Theory Division, CERN, CH-1211 Geneva 23, Switzerland}

\vspace{2cm} 
{\bf Abstract \\} 
\end{center} 
\setlength{\baselineskip}{0.3in} 

        We evaluate matching conditions for the Wilson coefficients of
operators mediating the $b \to s \gamma$ and $b \to s \; gluon$
transitions in a large class of extensions of the Standard Model. The
calculation is performed at the leading order in flavour-changing
couplings and includes two-loop QCD corrections. These corrections can
be numerically important when approximate cancellations occur
among the new physics contributions and/or the SM one. 

\vspace{1cm}

\setlength {\baselineskip}{0.2in}
\noindent \underline{\hspace{2in}}\\ 
\noindent
$^\star$ {\footnotesize Address in September 1997: Max Planck Institut
f{\"u}r Physik komplexer Systeme, N{\"o}thnitzerstr. 38,\\ D-01197 Dresden,
Germany. Permanent address: Institute of Theoretical Physics, Warsaw
University,\\ ul. Ho\.za 69, PL-00-681 Warsaw, Poland.}

\end{titlepage} 

\setlength{\baselineskip}{0.3in}

\noindent {\bf 1. Introduction}

        The decay \Bsg is known to be a sensitive test of various new
physics scenarios. Its branching ratio can significantly deviate from
the Standard Model prediction in the Supersymmetric Standard Model
(SSM) \cite{BBMR91,CDGG98}, multi-Higgs-doublet models
\cite{CDGG98,GSW90,BG98}, left--right-symmetric models
\cite{CM94,FY94,AI96} and other theories \cite{H93}. Since the
experimental results of CLEO \cite{CLEO98}
\be
BR[ B \to X_s \gamma] = (3.15 \pm 0.35_{stat} \pm 0.32_{syst} \pm 0.26_{model}) \times 10^{-4}
\ee
and ALEPH \cite{ALEPH98} 
\be
BR[ B \to X_s \gamma] = (3.11 \pm 0.80_{stat} \pm 0.72_{syst}) \times 10^{-4}
\ee
are consistent with the SM expectation (see below), parameter spaces
of these theories get severely constrained. However, given the present
sizable experimental errors, it is still conceivable that new physics
effects are large, but either tend to cancel in the decay rate or tend
to reverse the sign of the amplitude, when compared to the SM. In such
cases, next-to-leading QCD corrections to the new physics
contributions can be numerically very important \cite{CDGG98,BG98}.

        After the next-to-leading QCD corrections have been found
\cite{AG91}--\cite{CMM97} and confirmed \cite{P96}--\cite{BKP97} in the
Standard Model case, all the SM predictions given in the literature
fall in the range
\be
BR[ B \to X_s \gamma] = (3.3 \pm 0.3) \times 10^{-4}.
\ee
All the authors agree that the present theoretical uncertainty is
around 10\% (see e.g. \cite{CMM97}, \cite{BKP97}--\cite{GH98}).  Even
if all the authors had included the same set of corrections (which was
not the case), the central values of their predictions would be
allowed to differ by up to around 7\%, which is the expected size of
the uncalculated next-to-next-to-leading QCD corrections.

        Apart from the QCD corrections, two other important
contributions have been calculated recently, namely the $\Lambda/m_c$
non-perturbative corrections \cite{one.ov.mc} and the leading
electroweak corrections \cite{CM98,S98}. In both cases, the effects on
the branching ratio are below the $\;\sim \hspace{-1mm} 10\%$ overall
theoretical uncertainty, but the shifts in the central value are
relevant.\\

Theoretical analyses of $b \to s \gamma$ and $b \to s\;gluon$ are
usually performed in three steps. First, the full Standard Model (or
some of its extensions) is perturbatively matched on an effective
theory containing only light degrees of freedom, i.e. particles much
lighter than the electroweak gauge bosons. Flavour-changing
interactions in the low energy theory are mediated by effective
operators of dimension higher than four. In the second step, Wilson
coefficients of these operators are evolved with the help of the
renormalization group equations from the electroweak scale $\mu_0 \sim
M_W$ to the low-energy scale $\mu_b \sim m_b$. Finally, one evaluates
matrix elements of the effective operators between the physical states
of interest.

This procedure allows a resummation of large QCD logarithms such as
$[\al \ln(M_W^2/m_b^2)]^n$ from all orders of the perturbation series.
Moreover, its important advantage lies in that both the RGE evolution
and the calculation of matrix elements are practically the same in the
Standard Model and in many of its extensions. Thus, a variety of new
physics effects in \Bsg can be analysed with next-to-leading accuracy
without having to repeat the involved calculations of three-loop
anomalous dimensions \cite{CMM97,CMM98}, two-loop matrix elements
\cite{GHW96} or Bremsstrahlung corrections \cite{AG91,P96}. It is
sufficient to perform the matching calculation for $b \to s \gamma$
and $b \to s \; gluon$, including the potentially relevant
next-to-leading two-loop contributions.

Performing matching calculations in theories containing exotic
particles requires calculating basically the same (or at least very
similar) sets of Feynman diagrams as in the SM case. The variety of
final results is to a large extent due only to differences in
electromagnetic charges and QCD-transformation properties of the heavy
particles that are being decoupled. Consequently, it seems reasonable
to first calculate the matching conditions for a relatively generic
extension of the Standard Model. Results for various specific theories
can then be obtained by only substituting particular values of
couplings, charges and colour factors to the ``generic'' formulae.
Presenting such ``generic'' formulae for the one- and two-loop
matching conditions is the main purpose of the present paper.

The class of models we shall be interested in are theories in which
the leading contributions to $b \to s \gamma$ and $b \to s \; gluon$
originate from (heavy fermion)--(heavy boson) loops, similarly to the
SM case. Each such loop gives an additive contribution to the
considered Wilson coefficient. Thus, it is sufficient to perform a
calculation with only a single heavy fermion and a single heavy boson
(scalar or vector).  Having calculated the one-loop diagrams and
two-loop gluonic corrections to them, we shall be able to reproduce
the known next-to-leading matching results in the Standard
Model,\footnote{
except for the (light quark)--($W$-boson) contribution, which we calculate
separately.}
the Two-Higgs-Doublet Model (2HDM), and gluonic parts of the NLO
corrections to chargino contributions in the SSM. Examples of new
results that can be obtained from our formulae are the NLO matching
contributions in the left--right-symmetric models, as well as gluonic
parts of the NLO corrections to neutralino and gluino contributions in
the SSM.

        Completing the full NLO calculation in the SSM requires, in
addition, evaluating two-loop diagrams containing no gluons but only
heavy particles in internal lines. Some of those results are collected
in appendix C. The remaining contributions will be treated
approximately in the heavy gluino case.

Our paper is organized as follows. In the next section, we evaluate
leading and next-to-leading matching conditions originating from
(heavy fermion)--(heavy scalar) loops. In section~3, a similar
calculation is performed for (heavy fermion)--(heavy vector boson)
loops. Light-quark contributions are discussed in section~4. In
section~5, we specify what substitutions have to be made in the
general results, for particular extensions of the SM. Section~6 is
devoted to presenting a numerical example of the NLO matching
contribution effects. There, we consider the SSM with decoupled
gluino, which is relevant, for instance, in the context of models with
gauge-mediated SUSY breaking. Appendix A summarizes several useful
formulae which we have used for evaluating colour factors in the
generic case. Appendix B contains a list of functions that enter our
``generic'' results. Finally, appendix C contains results for matching
contributions originating from quartic squark vertices in the SSM.

\ \\
{\bf 2. Heavy fermion ~--~ heavy scalar loops}

        In the present section, we shall evaluate one- and two-loop
matching contributions for $b \to s \gamma$ and $b \to s \; gluon$ in
a theory described by the following lagrangian:
\bea 
{\cal L} &=& {\cal L}_{\scs QCD \times QED}(u,d,s,c,b)
+ (D^{\mu} \phi)^{\dagger} (D_{\mu} \phi) - m_{\phi}^2 \phi^{\dagger} \phi
+ \bar{\psi} ( i \slash D - m_{\psi}) \psi
\nonumber \\
&+& \left\{ C_{ijk} \phi^{\star}_i \bar{\psi}_j \left[ 
   (S_L P_L + S_R P_R) s_k + (B_L P_L + B_R P_R) b_k \right] + h.c. \right\}
+ {\cal L}_{\scs irrelevant}. \label{Lsf}
\eea
Here, ${\cal L}_{\scs QCD \times QED}(u,d,s,c,b)$ denotes kinetic
terms for the light quarks, photons and gluons as well as their gauge
interactions. The remainder of the first line contains kinetic, gauge
interaction and mass terms for a heavy complex scalar and a heavy
Dirac fermion. The covariant derivatives of these fields are the
following:
\bea
D_{\mu} \psi &=& \left[ \partial_{\mu} + i g_3 G_{\mu}^a T_{(\psi)}^a 
                                     + i e Q_{\psi} A_{\mu} \right] \psi,
\nonumber\\
D_{\mu} \phi &=& \left[ \partial_{\mu} + i g_3 G_{\mu}^a T_{(\phi)}^a 
                                     + i e Q_{\phi} A_{\mu} \right] \phi.
\eea

        The couplings $S_{L,R}$ and $B_{L,R}$ in eq.~(\ref{Lsf})
parametrize Yukawa interactions of the heavy particles with the $s$-
and $b$-quarks ($P_{L,R} = (1\mp\gamma_5)/2$). These interaction terms
must be QED- and QCD-singlets. The heavy particles can reside in any
representation of $SU(3)_{colour}$ for which such singlets exist. The
following gauge-invariance constraints must be satisfied by the
electric charges and colour generators:
\bea \label{constraints}
Q_{\psi} ~~~~~+~~~~~ Q_{\phi}~~~~ &=& -\f{1}{3}, \nonumber\\
T^a_{(\psi)jn} C_{ink} + T^a_{(\phi)in} C_{njk} &=& C_{ijn} T^a_{nk},
\eea
where $C_{ijk}$ are the Clebsch--Gordan coefficients contracting colour
indices in eq.~(\ref{Lsf}), and $T^a$ on the r.h.s. above is the
generator of the fundamental representation of $SU(3)$. All the
generators satisfy the standard commutation relations
\be
\left[ T^a_{(\psi)}, T^b_{(\psi)} \right] = i f_{abc} T^c_{(\psi)},
\hspace{2cm}
\left[ T^a_{(\phi)}, T^b_{(\phi)} \right] = i f_{abc} T^c_{(\phi)},
\hspace{2cm}
\left[ T^a, T^b \right] = i f_{abc} T^c.
\ee
The Casimir eigenvalues for the heavy particles will be denoted by
$\kappa_{\psi}$ and $\kappa_{\phi}$, respectively:
\be \label{kappa}
T^a_{(\psi)} T^a_{(\psi)} = \kappa_{\psi} \mbox{\large 1}, \hspace{3cm}
T^a_{(\phi)} T^a_{(\phi)} = \kappa_{\phi} \mbox{\large 1}.
\ee

        We assume that the Yukawa interactions $S_{L,R}$ and $B_{L,R}$
are weak, i.e. that it is mandatory to calculate at the leading order
in these interactions and $\alpha_{em}$, but at the next-to-leading
order in $\al$. Furthermore, we assume that all the remaining
interactions (denoted by ${\cal L}_{irrelevant}$) do not influence the
$b \to s \gamma$ and $b \to s \; gluon$ amplitudes at the considered
order in perturbation theory.

In each particular model, one needs to carefully verify whether the
latter assumption does not lead to neglecting important contributions.
Even in the Standard Model case, the top-quark Yukawa coupling $Y_t$
and the quartic Higgs coupling $\lambda$ can be similar in magnitude
to the QCD gauge coupling. Potentially, the ${\cal O}(Y_t,\lambda)$
corrections to the considered amplitudes could be almost as important
numerically as the ${\cal O}(\al)$ matching corrections. However, it
has been explicitly checked in ref.~\cite{S98} that this is not the
case for $b \to s \gamma$, i.e. that ${\cal O}(Y_t,\lambda)$ effects
in the SM are only around 1\% in the branching ratio, which is partly
due to accidental cancellations.

In the present and in the following three sections of this article, we
shall neglect ${\cal L}_{irrelevant}$, i.e. we shall restrict
ourselves only to the leading one-loop diagrams, and to
the next-to-leading two-loop diagrams {\em with gluons}.\footnote{
  In section 6, where the SSM case will be considered numerically, we
  shall include also flavour-conserving gluino couplings and the quartic
  squark couplings, which are proportional to $\al$.}
Such diagrams involving (heavy fermion)--(heavy scalar) loops are
presented in figs.~1 and 2, respectively. Small circles denote places
where the external gluon can couple in the $b \to s \; gluon$ case.
The external photon in the $b \to s \gamma$ case can couple in the
same places, except the ones denoted by ``5'' on internal gluon
propagators. Thus, we have 45 two-loop diagrams in the $b \to s \;
gluon$ case, and 37 two-loop diagrams in the $b \to s \gamma$ case.
\vspace{-3mm}
\begin{figure}[htbp]
\begin{center}
\epsfig{file=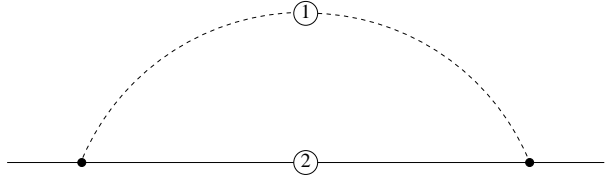,width=8cm}
\end{center}
\caption{One-loop diagrams}
\end{figure}
\begin{figure}[htbp]
\begin{center}
\begin{tabular}{p{0.48\linewidth}p{0.48\linewidth}}
\mbox{\epsfig{file=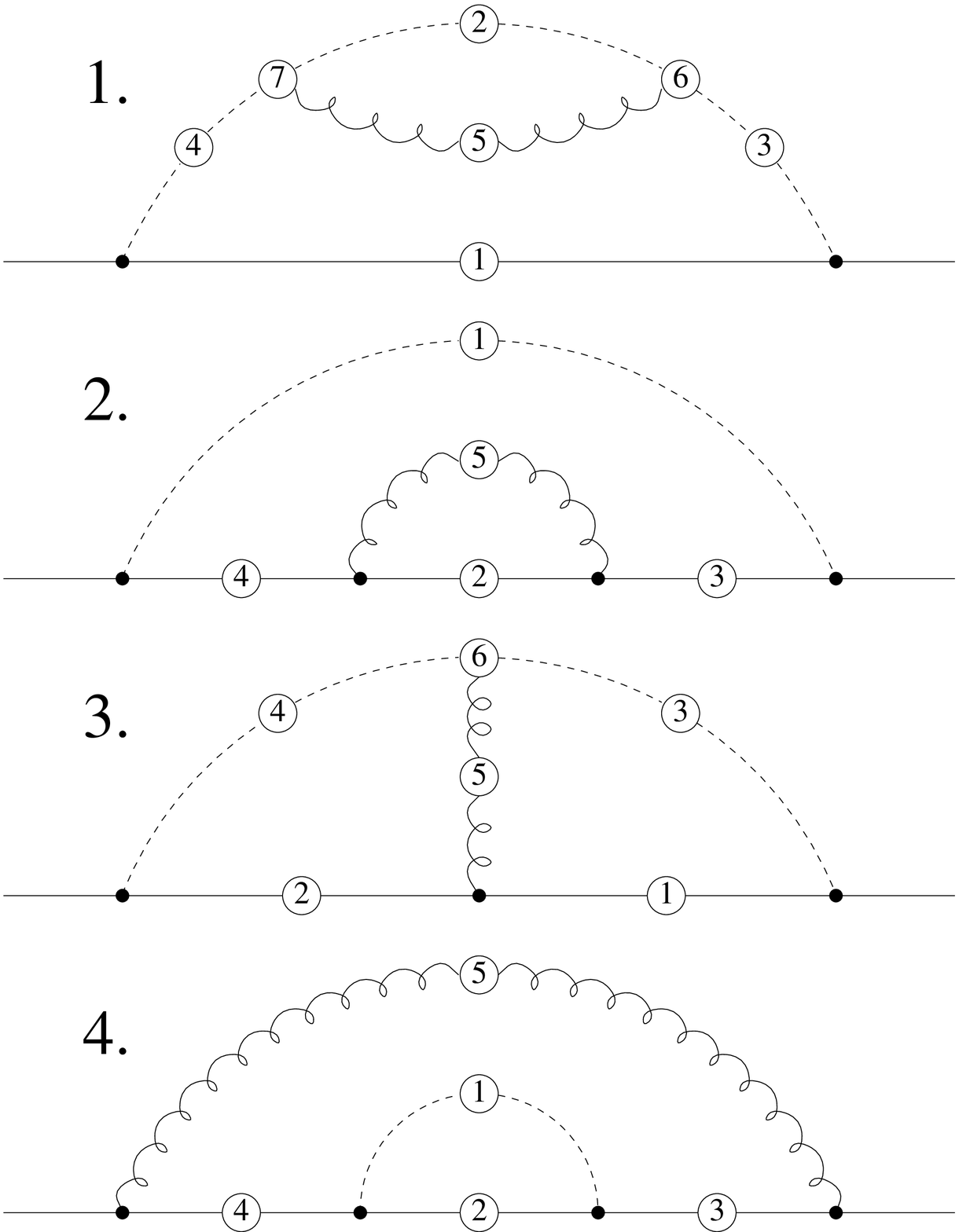,width=\linewidth}}  &
\mbox{\epsfig{file=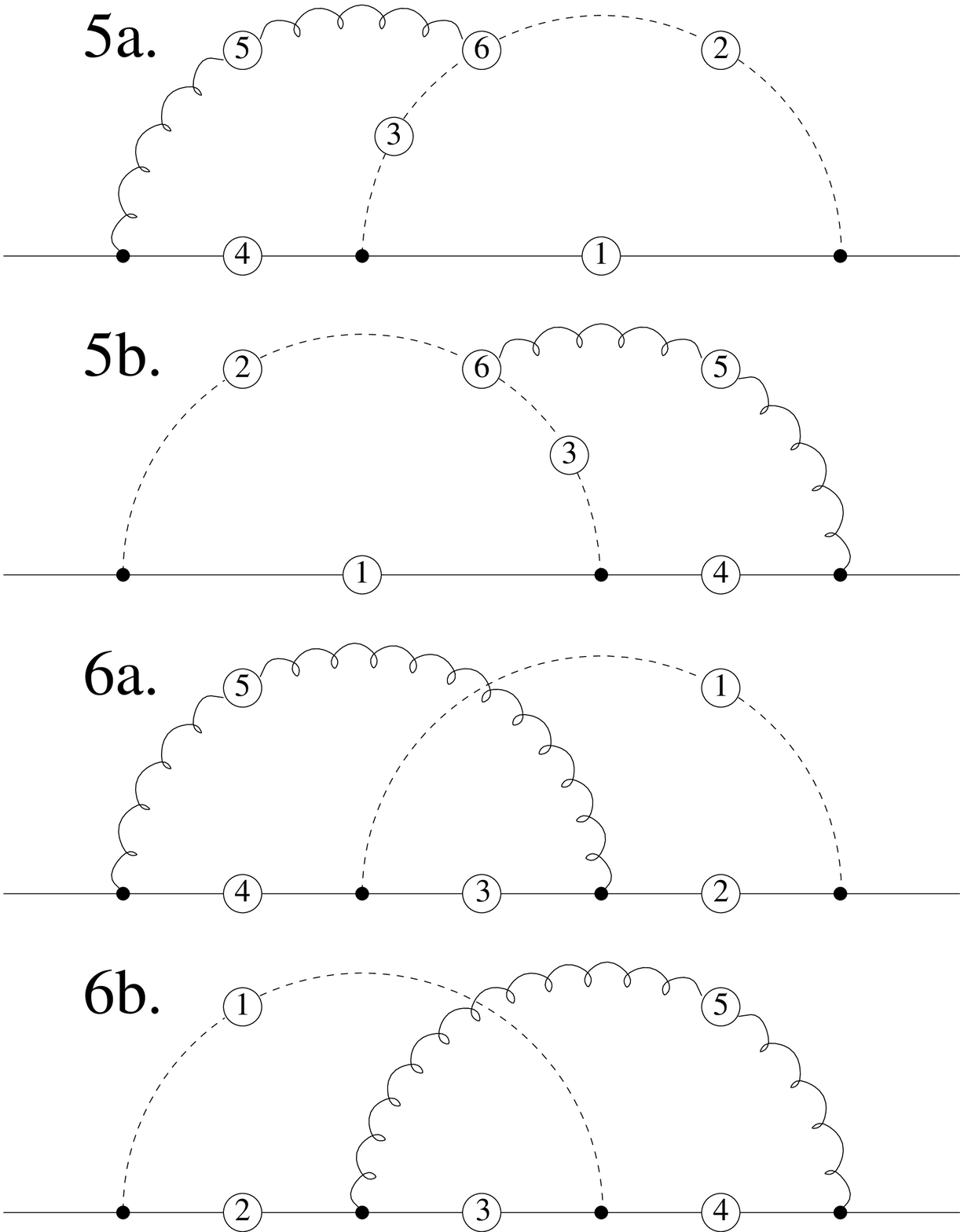,width=\linewidth}}
\end{tabular}
\end{center}
\caption{Two-loop diagrams}
\end{figure}

We evaluate all the diagrams off shell, in the 't~Hooft--Feynman
version of the background field gauge for QCD. For all the quantities
we need to renormalize, we use the $\overline{MS}$ scheme with the
renormalization scale $\mu_0$ that is assumed to be of the same order
as the heavy masses. Before performing the momentum integration, we
expand the integrands up to second order in (external momenta)/(heavy
masses). All the spurious IR divergences arising in this procedure are
regularized dimensionally.

        Next, we require equality of our result to the similar
off-shell 1PI Green function in the effective theory described by
the following effective lagrangian
\be \label{Leff} 
{\cal L}_{eff} = {\cal L}_{\scs QCD \times QED}(u,d,s,c,b)
\;+\; \f{4 G_F}{\sqrt{2}} V_{ts}^* V_{tb} \left[ 
\sum_{i=4,7,8} \left( C_i P_i + C'_i P'_i \right) \;+\; 
\left( \begin{array}{c} \mbox{\footnotesize EOM-vanishing}\\[-2mm] 
                        \mbox{\footnotesize   operators} \end{array} \right) \right],
\ee
where
\be \label{ope}
\begin{array}{rl}
P_4 = & (\bar{s} \gamma_{\mu} T^a P_L b) \sum_q (\bar{q}\gamma^{\mu} T^a q),      \\[0.2cm]
P_7  = &  \f{e}{16 \pi^2} m_b (\bar{s} \sigma^{\mu \nu} P_R b) F_{\mu\nu},        \\[0.2cm]
P_8  = &  \f{g_3}{16 \pi^2} m_b (\bar{s} \sigma^{\mu \nu} T^a P_R b) G_{\mu \nu}^a, \\[0.2cm]  
\end{array}
\ee
and the primed operators $P'_i$ are obtained from the ones above by
changing $P_{L,R}$ to $P_{R,L}$. The structure of the EOM-vanishing
operators (i.e. operators that vanish by the QCD$\times$QED
equations of motion) as well as other elements of our matching
procedure can be found in ref.~\cite{BMU98} where the Standard Model
case is described in detail.\footnote{
  Four-fermion operators generated by the photon exchange (such as
  $(\bar{s} \gamma_{\mu} P_L b) (\bar{e}\gamma^{\mu}e)$) are ignored
  here but included in ref.~\cite{BMU98}. They are irrelevant to $b
  \to s \gamma$ and $b \to s \; gluon$, but they matter, for instance,
  in $b \to s e^+ e^-$.}

        The obtained Wilson coefficients
\be
C_i(\mu_0) = C^{(0)}_i(\mu_0) + \f{\al(\mu_0)}{4 \pi} C^{(1)}_i(\mu_0) + ...
~~~~ (i=4,7,8)
\ee
(as well as their primed counterparts) can be written as linear
combinations of various functions of $x = (m_{\psi}/m_{\phi})^2$:
\bea
\begin{array}{lcrlr}
C_4^{(0)}(\mu_0) &=& 0,~~~~~ 
&&(\arabic{equation}) \addtocounter{equation}{1}\\[4mm]
C_7^{(0)}(\mu_0) &=& N \left\{ R_1 \right.
& \left[ g_1(x) Q_{\psi} - \f{1}{x} g_1\left(\f{1}{x}\right) Q_{\phi} \right] &\\[2mm]
&& + R_2 & \left.
\left[ g_2(x) Q_{\psi} + \left( x g_2(x) + \f{1}{2} \right) Q_{\phi} 
\right] \right\},
&(\arabic{equation}) \addtocounter{equation}{1}\\[4mm]
C_8^{(0)}(\mu_0) &=& N \left\{ R_1 \right.
& \left[ \f{3}{8}(\kappa_{\psi}-\kappa_{\phi}) 
  \left( g_1(x) + \f{1}{x} g_1\left(\f{1}{x}\right)\right) 
  + \f{1}{2} g_1(x) -\f{1}{2x} g_1\left(\f{1}{x}\right) \right] &\\[2mm] 
&& + R_2 & \left.
\left[ \f{3}{8}(\kappa_{\psi}-\kappa_{\phi}) \left((1-x)g_2(x) - \f{1}{2}\right) 
        + \f{x+1}{2} g_2(x) + \f{1}{4} \right] \right\},
&(\arabic{equation}) \addtocounter{equation}{1}\\[4mm]
C_4^{(1)}(\mu_0) &=& \multicolumn{2}{l}{ N S_L^* B_L 
\left[ (\kappa_{\phi} - \kappa_{\psi}) f_1(x) + f_2(x) \right],}
&(\arabic{equation}) \addtocounter{equation}{1}\\[4mm]
C_7^{(1)}(\mu_0) &=& 3 \ln \f{\mu_0^2}{m_{\phi}^2} &
\left[ (2 \kappa_{\psi} - \kappa_{\phi}) 
    x \left( \f{\partial C_7^{(0)}}{\partial x} \right)_{\hspace{-1.7mm}\scs S,B}
+ \left( \f{28}{9} + \kappa_{\psi} - \kappa_{\phi} \right) C_7^{(0)} 
+ ( \kappa_{\psi} - \f{4}{3} ) C_7^{(0){\scs LR}} - \f{16}{27} C_8^{(0)} \right]&\\[2mm] 
&& + N \left\{ R_1 \right. & \left[ 
  \left(  h_1(x) \kappa_{\psi} +   h_2(x) \kappa_{\phi} +   h_3(x) \right) Q_{\psi} 
+ \left(\;h_4(x) \kappa_{\psi} + \;h_5(x) \kappa_{\phi} + \;h_6(x) \right) Q_{\phi} 
\right] &\\[2mm] 
&& + R_2 & \left. \left[ 
  \left( h_7(x)    \kappa_{\psi} + h_8(x)    \kappa_{\phi} + h_9(x)    \right) Q_{\psi} 
+ \left( h_{10}(x) \kappa_{\psi} + h_{11}(x) \kappa_{\phi} + h_{12}(x) \right) Q_{\phi} 
\right] \right\}, 
&(\arabic{equation}) \addtocounter{equation}{1}\\[4mm]
C_8^{(1)}(\mu_0) &=& 3 \ln \f{\mu_0^2}{m_{\phi}^2} &
\left[ (2 \kappa_{\psi} - \kappa_{\phi}) 
    x \left( \f{\partial C_8^{(0)}}{\partial x} \right)_{\hspace{-1.7mm}\scs S,B}
+ \left( \f{26}{9} + \kappa_{\psi} - \kappa_{\phi} \right) C_8^{(0)} 
+ ( \kappa_{\psi} - \f{4}{3} ) C_8^{(0){\scs LR}} \right] &\\[4mm]
&& + N \left\{ R_1 \right. & \left[ 
  h_{13}(x) \kappa_{\psi}^2 + h_{14}(x) \kappa_{\phi}^2 + h_{15}(x) \kappa_{\psi} \kappa_{\phi} 
+ h_{16}(x) \kappa_{\psi}   + h_{17}(x) \kappa_{\phi}   + h_{18}(x) \right]\\[2mm] 
&& + R_2 & \left. \left[ 
  h_{19}(x) \kappa_{\psi}^2 + h_{20}(x) \kappa_{\phi}^2 + h_{21}(x) \kappa_{\psi} \kappa_{\phi} 
+ h_{22}(x) \kappa_{\psi}   + h_{23}(x) \kappa_{\phi}   + h_{24}(x) 
\right] \right\}
&(\arabic{equation}) \addtocounter{equation}{1}
\end{array}
\nonumber
\eea
with 
\be
R_1 = S_L^* B_L + \f{m_s}{m_b} S_R^* B_R 
\hspace{1cm} \mbox{and} \hspace{1cm} 
R_2 = S_L^* B_R \f{m_{\psi}}{m_b}.
\ee
The primed Wilson coefficients can be obtained from the ones above by
simply interchanging the chirality subscripts $L \leftrightarrow R$.
Explicit expressions for the functions $f_1(x)$, $f_2(x)$, $g_1(x)$,
$g_2(x)$ and $h_1(x)$--$h_{24}(x)$ are given in appendix B.

        The symbols $C_i^{(0){\scs LR}}$ used in the latter two
equations denote those parts of the leading-order coefficients
$C_i^{(0)}$ that are proportional to $S_L^* B_R$. The global
normalization constant $N$ equals to
\be 
N = \f{\xi \sqrt{2}}{8 m_{\phi}^2 G_F V_{ts}^* V_{tb}},
\ee
where $\xi$ parametrizes the contraction of the Clebsch--Gordan
coefficients:
\be \label{xi}
C_{ijk}^* C_{ijn} = \xi \delta_{kn}.
\ee

        The subscripts ``$S,B$'' at the derivatives of $C_7^{(0)}$ and
$C_8^{(0)}$ mean that the products $S^*_L B_L$, $S^*_R B_R$ and $S^*_L
B_R \f{m_{\psi}}{m_b}$ need to be treated as independent of $x$ when these
derivatives are taken. 

It is remarkable that colour factors of the two-loop diagrams could
have been reduced to quite a simple form, even though we have not
specified the representations of $SU(3)_{colour}$ in which the heavy
particles reside. Several identities that are useful in performing
this reduction are summarized in appendix A.

\ \\
{\bf 3. Heavy fermion ~--~ heavy vector boson loops}

        Here, we are going to consider only QCD-singlet vector bosons
$V_{\mu}$, and heavy fermions $\psi$ in the fundamental representation
of $SU(3)$. The lagrangian is now assumed to have the following form:
\bea
{\cal L} &=& {\cal L}_{\scs QCD \times QED}(u,d,s,c,b)
- (D_{\mu} V_{\nu})^* (D^{\mu} V^{\nu}) + m_V^2 V^*_{\mu} V^{\mu}
+ i e \omega V^*_{\mu} V_{\nu} F^{\mu\nu} + \bar{\psi} ( i \slash D - m_{\psi}) \psi
\nonumber \\
&+& \left\{ V^*_{\mu} \bar{\psi} \gamma^{\mu} \left[ 
   (\sigma_L P_L + \sigma_R P_R) s + (\beta_L P_L + \beta_R P_R) b \right] + h.c. \right\}
+ {\cal L}_{\scs irrelevant}, \label{Lvf}
\eea
where $F^{\mu\nu}$ is the electromagnetic field-strength tensor, and
\be
D_{\mu} V_{\nu} = \left( \partial_{\mu} + i e Q_{\scs V} A_{\mu}
\right) V_{\nu}
\mbox{~~~~~~with~~~~~~}
Q_{\scs V} = -Q_{\psi}-\f{1}{3}.
\ee
The constants $\sigma_{L,R}$ and $\beta_{L,R}$ are some arbitrary weak
coupling constants. If we substitute in eq.~(\ref{Lvf}) $V \to W$,
$\psi \to t$, $Q_{\scs V} \to -1$, $\omega \to -2$, $\sigma_R, \beta_R
\to 0$, $\sigma_L \to -\f{g_2}{\sqrt{2}} V_{ts}$ and $\beta_L \to
-\f{g_2}{\sqrt{2}} V_{tb}$, we obtain all the Standard Model
interactions that are relevant to the ($W$-boson)--(top quark) loop
contributions to $b \to s \gamma$ in the 't~Hooft--Feynman version of
the background field gauge.

The present calculation differs from the Standard Model one mainly by
that it allows right-handed couplings of the heavy vector boson to
fermions.  Such couplings occur for the $W$-boson in the
left--right-symmetric models. Their small magnitude can be compensated
by the large ratio $m_t/m_b$ in contributions to \Bsg
\cite{CM94,FY94}.

        The diagrams we need to consider now can be obtained from the
ones presented in figs.~1 and 2 by removing all the diagrams with
gluon--scalar couplings and replacing the scalar by the vector boson.
In this way, we obtain 16 two-loop diagrams in both the $b \to s \;
gluon$ and $b \to s \gamma$ cases.

        The effective lagrangian with which our full theory is being
matched remains the same as in eq.~(\ref{Leff}). The contributions to the
Wilson coefficients we obtain in the present case read
\bea
C_4^{(0)}(\mu_0) &=& 0,\\
C_7^{(0)}(\mu_0) &=& \tilde{N} \left[ 
\left( \sigma_L^* \beta_L + \f{m_s}{m_b} \sigma_R^* \beta_R \right) 
                      \left( j_1(x) Q_{\psi} + j_2(x) Q_V + j_3(x) \omega \right)
\right. \nonumber\\ && \left.
+ \sigma_L^* \beta_R \f{m_{\psi}}{m_b} 
                      \left( j_4(x) (Q_{\psi} + Q_V) + j_5(x) \omega \right) \right],\\
C_8^{(0)}(\mu_0) &=& \tilde{N} \left[ 
\left( \sigma_L^* \beta_L + \f{m_s}{m_b} \sigma_R^* \beta_R \right) j_1(x)
+ \sigma_L^* \beta_R \f{m_{\psi}}{m_b} j_4(x) \right],\\
C_4^{(1)}(\mu_0) &=& \tilde{N} \sigma_L^* \beta_L
\left[ \f{-12x^2+18x-4}{3(1-x)^4} \ln x + \f{-25x^2+29x+2}{9(1-x)^3} \right],\\
C_7^{(1)}(\mu_0) &=& \ln \f{\mu_0^2}{m_V^2} 
\left[ 8 x \left( \f{\partial C_7^{(0)}}{\partial x} \right)_{\hspace{-2mm}\sigma,\beta}
                       + \f{16}{3} C_7^{(0)} - \f{16}{9} C_8^{(0)} \right] 
\nonumber\\ && +\tilde{N} \left[ 
\left( \sigma_L^* \beta_L + \f{m_s}{m_b} \sigma_R^* \beta_R \right) 
                      \left( k_1(x) Q_{\psi} + k_2(x) Q_V + k_3(x) \omega \right)
\right. \nonumber\\ && \left.
+ \sigma_L^* \beta_R \f{m_{\psi}}{m_b} 
                  \left( k_4(x) Q_{\psi} + k_5(x) Q_V + k_6(x) \omega \right) \right],\\
C_8^{(1)}(\mu_0) &=& \ln \f{\mu_0^2}{m_V^2} 
\left[ 8 x \left( \f{\partial C_8^{(0)}}{\partial x} \right)_{\hspace{-2mm}\sigma,\beta}
                       + \f{14}{3} C_8^{(0)} \right] 
\nonumber\\ && +\tilde{N} \left[ 
\left( \sigma_L^* \beta_L + \f{m_s}{m_b} \sigma_R^* \beta_R \right) k_7(x)
+ \sigma_L^* \beta_R \f{m_{\psi}}{m_b} k_8(x) \right].
\eea
As before, contributions to the primed coefficients can be obtained by
simply interchanging the $L$ and $R$ subscripts in the couplings $\sigma$
and $\beta$. The normalization constant $\tilde{N}$ reads
\be 
\tilde{N} = \f{\sqrt{2}}{8 m_V^2 G_F V_{ts}^* V_{tb}}.
\ee
The functions $j_1(x)$--$j_5(x)$ and $k_1(x)$--$k_8(x)$ of
$x=(m_{\psi}/m_V)^2$ are given explicitly in appendix B.

\ \\
{\bf 4. Contributions from ($W$-boson)--(light quark) loops}

In order to obtain complete matching results, we need to consider
contributions from loops containing light quarks. In the Standard
Model case, loops with $W$-bosons and charm-quarks are relevant.
Diagrams with up-quarks are analogous, but numerically less important
because of the smallness of the CKM matrix element $V_{ub}$.

In the following, we shall assume that only the Standard Model
contributions are non-negligible in the light-quark case. This is a
correct assumption in many extensions of the Standard Model (e.g. 2HDM
or SSM) in which couplings of heavy scalars to light fermions are
suppressed by small Yukawa couplings. In the left--right-symmetric
models, there are additional contributions from right-handed coupling
of the $W$-boson to quarks. However, for the light quarks, such
contributions to $b \to s \gamma$ and $b \to s\;gluon$ are {\em not}
enhanced by the large ratio $m_t/m_b$. Consequently, they are
negligible when compared to the top-quark one.\footnote{
  Unless the left- and right-handed CKM matrices have a very different
  hierarchy of their elements, which we assume not to be the case
  here.}

        The Standard Model diagrams we need to consider here are the
same as in the previous section, with the heavy fermion replaced by
the light quark. However, the effective theory side is now somewhat
more complicated, since we need to include operators that contain
light quarks. In the charm-quark case, they read
\be \label{ope.light}
\begin{array}{rl}
        P_1 = & (\bar{s} \gamma_{\mu} P_L T^a c) (\bar{c} \gamma^{\mu} P_L T^a b), \\[0.2cm]
        P_2 = & (\bar{s} \gamma_{\mu} P_L     c) (\bar{c} \gamma^{\mu} P_L     b). 
\end{array}
\ee
Their Wilson coefficients are found by considering the
$\bar{s}b\bar{c}c$ Green function\footnote{
See ref.~\cite{BMU98} for more details.}
\be
\begin{array}{ll}
C_1^{(0)}(\mu_0) = 0, \hspace{2cm}&
C_2^{(0)}(\mu_0) = r_{ct},\\[3mm]
C_1^{(1)}(\mu_0) = \left[ -15 - 6 \ln \f{\mu_0^2}{M_W^2}\right] r_{ct}, \hspace{2cm}&
C_2^{(1)}(\mu_0) = 0.
\end{array}
\ee
where $r_{ct} = V_{cs}^* V_{cb} / V_{ts}^* V_{tb}$. The value of
$C_1^{(1)}$ has been obtained in the $\overline{MS}$ scheme (applied
throughout this paper) and using the evanescent operator
\be
E_1 = (\bar{s} \gamma_{\mu} \gamma_{\nu} \gamma_{\rho} P_L T^a c) 
      (\bar{c} \gamma^{\mu} \gamma^{\nu} \gamma^{\rho} P_L T^a b) - 16 P_1.
\ee

        After performing the matching for $P_1$ and $P_2$,
contributions to the Wilson coefficients of $P_4$, $P_7$ and $P_8$ can be
found. We obtain
\bea
\begin{array}{lllr}
\hspace{1cm} &
\delta^c C_4^{(0)}(\mu_0) = 0, \hspace{4cm} &  
\delta^c C_4^{(1)}(\mu_0) = \left(    \f{7}{9}   - \f{2}{3}  \ln \f{\mu_0^2}{M_W^2} \right) r_{ct},
& \hspace{1.2cm} (\arabic{equation}) \addtocounter{equation}{1}\\[3mm]
\end{array} \nonumber
\eea
\bea
\begin{array}{lllr}
\hspace{0.5cm} &
\delta^c C_7^{(0)}(\mu_0) = \f{23}{36} r_{ct}, \hspace{3cm} & 
\delta^c C_7^{(1)}(\mu_0) = \left( -\f{713}{243} - \f{4}{81} \ln \f{\mu_0^2}{M_W^2} \right) r_{ct},
& \hspace{1.4cm} (\arabic{equation}) \addtocounter{equation}{1}\\[3mm]
& 
\delta^c C_8^{(0)}(\mu_0) = \f{1}{3} r_{ct}, &
\delta^c C_8^{(1)}(\mu_0) = \left( -\f{91}{324}  + \f{4}{27} \ln \f{\mu_0^2}{M_W^2} \right) r_{ct}.
& (\arabic{equation}) \addtocounter{equation}{1}
\end{array} \nonumber
\eea
Results for the up-quark loops are obtained from the above ones by
replacing $r_{ct}$ by \linebreak $r_{ut} = V_{us}^* V_{ub} / V_{ts}^*
V_{tb}$.  After adding the up- and charm-quark contributions, we can
make use of the equality ~$r_{ut} + r_{ct} = -1$,~ which follows from
unitarity of the CKM matrix.

        The primed coefficients are negligible in the Standard Model
case, because they are proportional to small Yukawa couplings of the
would-be Goldstone boson to light quarks.

\ \\
{\bf 5. Substitutions}

        Let us first use our generic results to reproduce the known SM
matching conditions at the next-to-leading order in QCD. The only
scalar we need to consider in the SM is the charged would-be Goldstone
boson. We use the 't~Hooft--Feynman version of the background field
gauge, in which the ($W$-boson)-photon-scalar vertices are absent. The
only heavy fermion is the top quark, for which one finds
\bea
N \left( S_L^* B_L + \f{m_s}{m_b} S_R^* B_R \right) &=& 
\f{x}{2} \left( 1 + \f{m_s^2}{M_W^2} \right) \simeq \f{x}{2}
\hspace{2cm} \mbox{with~~} x = \f{m_t^2}{M_W^2},\\
N S_L^* B_R \f{m_{\psi}}{m_b} &=& -\f{x}{2},\\
\tilde{N} \sigma_L^* \beta_L &=& \f{1}{2}
\hspace{4.5cm} \mbox{and~~} \sigma_R = \beta_R = 0.
\eea
The remaining substitutions one needs to make are 
$m_{\psi} \to m_t$,~ 
$m_{\phi},m_V \to M_W$,~
$Q_{\psi} \to \f{2}{3}$,~ 
$Q_{\phi},Q_V \to -1$,~
$\omega \to -2$,~
$\kappa_{\psi} \to \f{4}{3}$~
and $\kappa_{\phi} \to 0$. 
Adding up the results from sections 2, 3 and 4, one easily finds 
\bea 
C_{7,8}^{(0)}(\mu_0) =  -\f{3}{2} x H_1^{[7,8]}(x) \hspace{1cm} \mbox{with~~~~~}
\left\{ \begin{array}{lclr}
H_1^{[7]}(x) &=& \f{-3x^2+2x}{6(1-x)^4}\ln x \;+\; \f{-8x^2-5x+7}{36(1-x)^3}\\
&&& \hspace{2.5cm} (\arabic{equation}) \addtocounter{equation}{1}\\
H_1^{[8]}(x) &=& \f{x}{2(1-x)^4} \ln x \;+\; \f{-x^2+5x+2}{12(1-x)^3}
\end{array} \right. \nonumber \\[2mm]
\begin{array}{lclr}
C_4^{(1)}(\mu_0) &=&  \f{4x^4-16x^3+15x^2}{6(1-x)^4} \ln x \;+\; \f{7x^3-35x^2+42x-8}{12(1-x)^3}
                \;+\; \f{2}{3} \ln \f{\mu_0^2}{m_t^2},
&(\arabic{equation}) \addtocounter{equation}{1}\\[4mm]
C_7^{(1)}(\mu_0) &=&  \f{-16x^4-122x^3+80x^2-8x}{9(1-x)^4} Li_2\left(1-\f{1}{x}\right) 
                \;+\; \f{-4x^5+407x^4+1373x^3-957x^2+45x}{81(1-x)^5} \ln x \\[2mm]  
&& \hspace{-2cm}  +\; \f{1520x^4+12961x^3-12126x^2+3409x-580}{486(1-x)^4}
                \;+\; \left[ \f{6x^4+46x^3-28x^2}{3(1-x)^5} \ln x 
\;+\; \f{106x^4+287x^3+1230x^2-1207x+232}{81(1-x)^4} \right] \ln \f{\mu_0^2}{m_t^2},
&(\arabic{equation}) \addtocounter{equation}{1}\\[4mm]
C_8^{(1)}(\mu_0) &=&  \f{-4x^4+40x^3+41x^2+x}{6(1-x)^4} Li_2\left(1-\f{1}{x}\right) 
                \;+\; \f{32x^5-16x^4-2857x^3-3981x^2-90x}{216(1-x)^5} \ln x \\[2mm] 
&& \hspace{-2cm}  +\;  \f{611x^4-13346x^3-29595x^2+1510x-652}{1296(1-x)^4} 
                \;+\; \left[ \f{-17x^3-31x^2}{2(1-x)^5} \ln x
\;+\; \f{89x^4-446x^3-1437x^2-950x+152}{108(1-x)^4} \right] \ln \f{\mu_0^2}{m_t^2}.
&(\arabic{equation}) \addtocounter{equation}{1} \end{array} \nonumber
\eea
These results agree with the main findings of
refs.~\cite{AY94,GH97,BKP97}.  Contributions to the primed
coefficients can be neglected in the SM, because they are suppressed
by $m_s/m_b$.

        Let us now turn to the Two-Higgs-Doublet Model II. For
diagrams with the physical charged higgson exchanges, the necessary
substitutions in the formulae of section 2 are the following:
$m_{\psi} \to m_t$,~ 
$m_{\phi} \to M_{H^-}$,~
$Q_{\psi} \to \f{2}{3}$,~ 
$Q_{\phi} \to -1$,~
$\kappa_{\psi} \to \f{4}{3}$,~
$\kappa_{\phi} \to 0$~ and
\bea
N \left( S_L^* B_L + \f{m_s}{m_b} S_R^* B_R \right) &=& 
\f{y}{2} \cot^2 \beta \left( 1 + \f{m_s^2}{M_W^2} \right) \simeq \f{y}{2} \cot^2 \beta,\\
N S_L^* B_R \f{m_{\psi}}{m_b} &=& \f{y}{2}
\hspace{5cm} \mbox{with~~} y = \f{m_t^2}{M_{H^-}^2}.
\eea
The resulting contributions to the Wilson coefficients read
\bea
\delta^{\scs H^-} C_{7,8}^{(0)}(\mu_0) = -\f{y \cot^2 \beta}{2} H_1^{[7,8]}(y) 
                                   + \f{1}{2} H_2^{[7,8]}(y)
\mbox{~~~~~with~~}
\left\{ \begin{array}{lclr}
H_2^{[7]}(y) &=& \f{-3y^2+2y}{3(1-y)^3} \ln y \;+\; \f{-5y^2+3y}{6(1-y)^2}\\
&&&(\arabic{equation}) \addtocounter{equation}{1}\\
H_2^{[8]}(y) &=& \f{y}{(1-y)^3} \ln y \;+\; \f{-y^2+3y}{2(1-y)^2}
\end{array} \right. \nonumber\\
\begin{array}{lclr}
\delta^{\scs H^-} C_4^{(1)}(\mu_0) &=&  \cot^2 \beta \left[ 
         \f{3y^2-2y}{6(1-y)^4} \ln y \;+\; \f{-7y^3+29y^2-16y}{36(1-y)^3} \right],
&(\arabic{equation}) \addtocounter{equation}{1}\\[4mm]
\delta^{\scs H^-} C_7^{(1)}(\mu_0) &=&  \cot^2 \beta \left\{ 
      \f{16y^4-74y^3+36y^2}{9(1-y)^4} Li_2\left(1-\f{1}{y}\right) \right. 
\;+\;  \f{-63y^4+807y^3-463y^2+7y}{81(1-y)^5} \ln y \\[2mm]
 &+& \f{-1202y^4+7569y^3-5436y^2+797y}{486(1-y)^4} 
\;+\; \left. \left[ \f{6y^4+46y^3-28y^2}{9(1-y)^5} \ln y 
      \;+\; \f{-14y^4+135y^3-18y^2-31y}{27(1-y)^4}  \right]  \ln \f{\mu_0^2}{m_t^2}\right\}\\[2mm]
 &+& \f{-32y^3+112y^2-48y}{9(1-y)^3} Li_2\left(1-\f{1}{y}\right) 
\;+\;  \f{14y^3-128y^2+66y}{9(1-y)^4} \ln y 
\;+\; \f{8y^3-52y^2+28y}{3(1-y)^3} \\[2mm] 
 &+& \left[ \f{-12y^3-56y^2+32y}{9(1-y)^4} \ln y 
      \;+\; \f{16y^3-94y^2+42y}{9(1-y)^3} \right]  \ln \f{\mu_0^2}{m_t^2},
&(\arabic{equation}) \addtocounter{equation}{1}\\[4mm]
\delta^{\scs H^-} C_8^{(1)}(\mu_0) &=& \cot^2 \beta \left\{ 
      \f{13y^4-17y^3+30y^2}{6(1-y)^4} Li_2\left(1-\f{1}{y}\right) \right. 
\;+\;   \f{-468y^4+321y^3-2155y^2-2y}{216(1-y)^5} \ln y \\[2mm]
 &+& \f{-4451y^4+7650y^3-18153y^2+1130y}{1296(1-y)^4} 
\;+\; \left. \left[ \f{-17y^3-31y^2}{6(1-y)^5} \ln y 
      \;+\; \f{-7y^4+18y^3-261y^2-38y}{36(1-y)^4}  \right]  \ln \f{\mu_0^2}{m_t^2}\right\}\\[2mm]
 &+&  \f{-17y^3+25y^2-36y}{6(1-y)^3} Li_2\left(1-\f{1}{y}\right) 
\;+\; \f{34y^3-7y^2+165y}{12(1-y)^4} \ln y 
\;+\; \f{29y^3-44y^2+143y}{8(1-y)^3} \\[2mm]
 &+& \left[ \f{17y^2+19y}{3(1-y)^4} \ln y 
      \;+\; \f{7y^3-16y^2+81y}{6(1-y)^3} \right]  \ln \f{\mu_0^2}{m_t^2}.
&(\arabic{equation}) \addtocounter{equation}{1} \end{array} \nonumber
\eea
These results are in agreement with refs.~\cite{CDGG98,BG98}.
Similarly to the SM case, the primed coefficients are suppressed by
$m_s/m_b$, and can be neglected.\\

        The third example we would like to discuss here is the
$SU(2)_L \times SU(2)_R \times U(1)$ model. The notation introduced in
ref.~\cite{CM94} will be followed. We shall restrict ourselves to such
contributions from the light $W$-boson (and the corresponding would-be
Goldstone boson) which are suppressed by the small $W$-boson mixing angle
$\zeta$, but simultaneously enhanced by the large quark-mass ratio
$\f{m_t}{m_b}$. In this case, the substitutions one needs to make in
the results of sections 2 and 3 are the following:
\be
\begin{array}{ll}
N S_L^* B_R \f{m_{\psi}}{m_b} = \f{x}{2} A^{tb}, \hspace{3cm}& 
\tilde{N} \sigma_L^* \beta_R \f{m_{\psi}}{m_b} = \f{1}{2} A^{tb},\\[2mm]
N S_R^* B_L \f{m_{\psi}}{m_b} = \f{x}{2} \left(A^{ts}\right)^*, & 
\tilde{N} \sigma_R^* \beta_L \f{m_{\psi}}{m_b} = \f{1}{2} \left(A^{ts}\right)^*,
\end{array}
\ee
where $x = \f{m_t^2}{M_W^2}$ and
\be
A^{tb(s)} = \f{m_t}{m_b} \; \zeta e^{i\alpha} \; 
\f{g_{2R} V^R_{tb(s)}}{g_{2L} V^L_{tb(s)}} \;\;+\;\; {\cal O}(\zeta^2).
\ee
The ``left--left'' and ``right--right'' products of couplings can be
set to zero now, because their non-SM parts are of order ${\cal
  O}(\zeta^2,\f{m_b}{M_W})$, i.e. negligibly small. Substitutions for
masses, charges and colour factors here are the same as for top-quark
loops in the Standard Model case.

        The obtained contributions to the Wilson coefficients read
\bea
\begin{array}{lclr}
\delta^{\scs LR} C_4(\mu_0) &=&  0,
&(\arabic{equation}) \addtocounter{equation}{1}\\[2mm]
\delta^{\scs LR} C_7^{(0)}(\mu_0) &=& A^{tb} \left[ \f{3x^2-2x}{2(1-x)^3} \ln x 
                             \;+\; \f{-5x^2+31x-20}{12(1-x)^2} \right],
&(\arabic{equation}) \addtocounter{equation}{1}\\[4mm]
\delta^{\scs LR} C_8^{(0)}(\mu_0) &=& A^{tb} \left[ \f{-3x}{2(1-x)^3} \ln x 
                             \;+\; \f{-x^2-x-4}{4(1-x)^2} \right],
&(\arabic{equation}) \addtocounter{equation}{1}\\[4mm]
\delta^{\scs LR} C_7^{(1)}(\mu_0) &=&  A^{tb} \left\{
                      \f{-32x^3-112x^2+48x}{9(1-x)^3}  Li_2\left(1-\f{1}{x}\right) 
                \;+\; \f{86x^3+120x^2-30x-32}{9(1 - x)^4} \ln x \right. \\[2mm]
                 &+& \left. \f{24x^3+320x^2-220x+20}{9(1-x)^3} 
                \;+\; \left[ \f{12x^3+56x^2-32x}{3(1-x)^4} \ln x 
                \;+\;  \f{16x^3+90x^2+66x-64}{9(1-x)^3}  \right] \ln \f{\mu_0^2}{m_t^2} \right\},
&(\arabic{equation}) \addtocounter{equation}{1}\\[4mm]
\delta^{\scs LR} C_8^{(1)}(\mu_0) &=&  A^{tb} \left\{
                      \f{-17x^3+89x^2+12x}{6(1-x)^3} Li_2\left(1-\f{1}{x}\right) 
                \;+\;  \f{34x^3-375x^2-207x-28}{12(1 - x)^4} \ln x \right. \\[2mm]
                 &+& \left. \f{87x^3-640x^2-451x-148}{24(1-x)^3} 
                \;+\; \left[ \f{-17x^2-19x}{(1 - x)^4} \ln x
                \;+\; \f{7x^3-36x^2-159x-28}{6(1 - x)^3} \right] \ln \f{\mu_0^2}{m_t^2} \right\}.
&(\arabic{equation}) \addtocounter{equation}{1} \end{array} \nonumber
\eea
The primed coefficients are obtained from the above ones by changing
$A^{tb}$ to $\left( A^{ts} \right)^*$. Expressions for the
leading-order coefficients are in agreement with ref.~\cite{CM94}. The
next-to-leading results are new.

        Physical charged scalars present in the left--right-symmetric
models can give contributions enhanced by $\f{m_t}{m_b}$, too
\cite{AI96}. Such contributions can be calculated with the help of our
generic formulae as well. However, their explicit form and potential
numerical importance depend on details of the Higgs sector that is
not unique in these models.\\

        Finally, let us turn to the Supersymmetric Standard Model. In
this case, we shall {\em exactly} follow the notation of
ref.~\cite{CMW96}.\footnote{
The CKM matrix will be denoted by $K$ in all the SSM formulae.}
In particular, the relevant couplings of charginos, neutralinos and
gluinos to matter will be parametrized by $X^{U_{L,R}}$, $Z^{D_{L,R}}$
and $g_3 \Gamma^{D_{L,R}}$, respectively.  However, we shall formally
treat all these matrices as independent from other parameters of the
model. Only diagrams containing gluons have been included in their
one-loop QCD renormalization, and we have used the $\overline{MS}$
scheme in dimensional regularization (not in dimensional
reduction). In effect, immediate substitution of tree-level
supersymmetric expressions for these matrices (eq.~(2.12) of
ref.~\cite{CMW96}) is allowed only for the leading-order terms in the
results presented below. Section 6 contains a discussion of
applicability and treatment of the ${\cal O}(\al)$ terms.

In the results presented below, flavour-conserving gluino interactions
as well as strong quartic squark interactions have been formally
treated as the weak ones. Additional contributions to the considered
Wilson coefficients from two-loop diagrams with strong quartic squark
interactions are collected in appendix C. A complete inclusion of
strong gluino coupling effects at NLO is beyond the scope of the
present paper. Consequently, our results can be used for making
numerical predictions in the SSM only in the heavy-gluino limit (see
section 6).

At one loop in the SSM, $b \to s \gamma$ and $b \to s \;gluon$
off-shell Green functions receive contributions from
chargino--squark, neutralino--squark and gluino--squark loops, in
addition to the SM and charged higgson contributions that have been
already discussed. For the loops containing chargino
$\tilde{\chi}^-_{\scs I}$~ ($I=1,2$) and the up-squark $\tilde{u}_A$~
($A=1,...,6$), as well as for the gluonic corrections to them, we need
to make the following substitutions in section 2:
$m_{\psi} \to m_{\tilde{\chi}^-_I}$,~ 
$m_{\phi} \to m_{\tilde{u}_A}$,~
$Q_{\psi} \to -1$,~ 
$Q_{\phi} \to 2/3$,~
$\kappa_{\psi} \to 0$,~
$\kappa_{\phi} \to 4/3$~ and
\bea
N \left( S_L^* B_L + \f{m_s}{m_b} S_R^* B_R \right) &=& 
\f{1}{g_2^2 K_{ts}^* K_{tb}} \; \f{M_W^2}{m_{\tilde{u}_A}^2} \;
\left( X^{U_L}_I \right)^*_{A2} \left( X^{U_L}_I \right)_{A3}
\;\;+\;\; {\cal O}\left(\f{m_s^2}{M_W^2}\right),
\nonumber\\
N S_L^* B_R \f{m_{\psi}}{m_b} &=& 
\f{1}{g_2^2 K_{ts}^* K_{tb}} \; \f{M_W^2}{m_{\tilde{u}_A}^2} \;
\left( X^{U_L}_I \right)^*_{A2} \left( X^{U_R}_I \right)_{A3}
\f{m_{\tilde{\chi}^-_I}}{m_b}.
\eea
After summing up all the chargino and up-squark species, we obtain the
following contributions to the Wilson coefficients 
\bea \label{chargino4}
\delta^{\tilde{\chi}^-} C_4^{(1)}(\mu_0) &=&  
\f{1}{g_2^2 K_{ts}^* K_{tb}} \sum_{A=1}^6 \sum_{I=1}^2 \f{M_W^2}{m_{\tilde{\chi}^-_I}^2} 
        \left( X^{U_L}_I \right)^*_{A2} \left( X^{U_L}_I \right)_{A3} H_1^{[4]}(z_{\scs AI}) 
\hspace{1.3cm} \mbox{with~~} z_{\scs AI}= \f{m_{\tilde{u}_A}^2}{m_{\tilde{\chi}^-_I}^2},\\[2mm]
\delta^{\tilde{\chi}^-} C_{7,8}(\mu_0) &=&  
\f{1}{g_2^2 K_{ts}^* K_{tb}} \sum_{A=1}^6 \sum_{I=1}^2 \f{M_W^2}{m_{\tilde{\chi}^-_I}^2} 
\times \nonumber\\ && \hspace{-1.8cm} \times \left\{
  \left( X^{U_L}_I \right)^*_{A2} \left( X^{U_L}_I \right)_{A3}
                \left[ H_1^{[7,8]}  (z_{\scs AI}) \hspace{1.3cm}
+ \f{\al}{4\pi} \left( H_1^{[7,8]'} (z_{\scs AI}) 
                     + H_1^{[7,8]''}(z_{\scs AI}) 
    \ln \left( \f{\mu_0^2}{m_{\tilde{u}_A}^2}\right) \right) \right]
\right. \nonumber\\ && \hspace{-2.5cm}
+ \left. \f{m_{\tilde{\chi}^-_I}}{m_b} \left( X^{U_L}_I \right)^*_{A2} \left( X^{U_R}_I \right)_{A3} 
                \left[ H_2^{[7,8]}  (z_{\scs AI}) + \lambda^{[7,8]} 
+ \f{\al}{4\pi} \left( H_2^{[7,8]'} (z_{\scs AI}) 
                     + H_2^{[7,8]''}(z_{\scs AI}) 
    \ln \left( \f{\mu_0^2}{m_{\tilde{u}_A}^2}\right) \right) \right] \right\}. \hspace{1cm} 
\label{chargino78}
\eea
The constants $\lambda^{[7]}=\f{5}{6}$ and $\lambda^{[8]}=\f{1}{2}$
drop out by unitarity of squark mixing matrices when the
chargino--matter couplings $X^{U_{L,R}}$ are no longer treated as
arbitrary but expressed in terms of other SSM parameters. The
functions $H_{1,2}^{[7,8]}$ have already been encountered in the SM
and 2HDM cases. The functions $H_1^{[4]}$, $H_{1,2}^{[7,8]'}$ and
$H_{1,2}^{[7,8]''}$ have the following explicit form:
\bea
\begin{array}{lclr}
H_1^{[4]}(x) &=& \f{-1}{3(1-x)^4} \ln x \;+\; \f{-2x^2+7x-11}{18(1-x)^3},
&(\arabic{equation}) \addtocounter{equation}{1}\\[4mm]
H_1^{[7]'}(x)  &=&  \f{24x^3+52x^2-32x}{9(1-x)^4} Li_2\left(1-\f{1}{x}\right) 
              \;+\; \f{-189x^3-783x^2+425x+43}{81(1-x)^5} \ln x 
              \;+\; \f{-1030x^3-1899x^2+1332x+85}{243(1-x)^4},
&(\arabic{equation}) \addtocounter{equation}{1}\\[4mm]
H_1^{[7]''}(x) &=&  \f{6x^3-62x^2+32x}{9(1-x)^5} \ln x
              \;+\; \f{28x^3-129x^2-12x+41}{27(1-x)^4},
&(\arabic{equation}) \addtocounter{equation}{1}\\[4mm]
H_2^{[7]'}(x)  &=&   \f{112x^2-48x}{9(1-x)^3} Li_2\left(1-\f{1}{x}\right)
              \;+\;   \f{12x^3-176x^2+64x+16}{9(1-x)^4} \ln x 
              \;+\;  \f{-170x^2+66x+20}{9(1-x)^3},
&(\arabic{equation}) \addtocounter{equation}{1}\\[4mm]
H_2^{[7]''}(x) &=&   \f{12x^3-88x^2+40x}{9(1-x)^4} \ln x
              \;+\;  \f{-14x^2-54x+32}{9(1-x)^3},
&(\arabic{equation}) \addtocounter{equation}{1}\\[4mm]
H_1^{[8]'}(x)  &=&  \f{-9x^3-46x^2-49x}{12(1-x)^4} Li_2\left(1-\f{1}{x}\right) 
              \;+\;  \f{81x^3+594x^2+1270x+71}{108(1-x)^5} \ln x 
              \;+\;  \f{923x^3+3042x^2+6921x+1210}{648(1-x)^4},
&(\arabic{equation}) \addtocounter{equation}{1}\\[4mm]
H_1^{[8]''}(x) &=&  \f{5x^2+19x}{3(1-x)^5} \ln x
              \;+\; \f{7x^3-30x^2+141x+26}{18(1-x)^4},
&(\arabic{equation}) \addtocounter{equation}{1}\\[4mm]
H_2^{[8]'}(x)  &=&   \f{-16x^2-12x}{3(1-x)^3} Li_2\left(1-\f{1}{x}\right) 
              \;+\;  \f{52x^2+109x+7}{6(1-x)^4} \ln x 
              \;+\;  \f{95x^2+180x+61}{12(1-x)^3},
&(\arabic{equation}) \addtocounter{equation}{1}\\[4mm]
H_2^{[8]''}(x) &=&   \f{10x^2+26x}{3(1-x)^4} \ln x
              \;+\;  \f{-x^2+30x+7}{3(1-x)^3}.
\label{chargino.end}
&(\arabic{equation}) \addtocounter{equation}{1} \end{array} \nonumber
\eea
Contributions to the primed coefficients can be obtained by
interchanging $X^{U_L}$ and $X^{U_R}$ in eqs.~(\ref{chargino4}) and
(\ref{chargino78}). Their suppression by $m_s/m_b$ becomes transparent
only after expressing $X^{U_R}$ in terms of other SSM parameters (cf.
eq.~(\ref{xur}) below).\\

Very similar substitutions need to be made for loops containing the
neutralino $\tilde{\chi}^0_{\scs I}$~ ($I=1,2,3,4$) and the
down squark $\tilde{d}_A$~ ($A=1,...,6$):
$m_{\psi} \to m_{\tilde{\chi}^0_I}$,~ 
$m_{\phi} \to m_{\tilde{d}_A}$,~
$Q_{\psi} \to 0$,~ 
$Q_{\phi} \to -1/3$,~
$\kappa_{\psi} \to 0$,~
$\kappa_{\phi} \to 4/3$~ and
\bea
N \left( S_L^* B_L + \f{m_s}{m_b} S_R^* B_R \right) &=& 
\f{1}{g_2^2 K_{ts}^* K_{tb}} \; \f{M_W^2}{m_{\tilde{d}_A}^2} \;
\left( Z^{D_L}_I \right)^*_{A2} \left( Z^{D_L}_I \right)_{A3}
\;\;+\;\; {\cal O}\left(\f{m_s^2}{M_W^2}\right),
\nonumber\\
N S_L^* B_R \f{m_{\psi}}{m_b} &=& 
\f{1}{g_2^2 K_{ts}^* K_{tb}} \; \f{M_W^2}{m_{\tilde{d}_A}^2} \;
\left( Z^{D_L}_I \right)^*_{A2} \left( Z^{D_R}_I \right)_{A3}
\f{m_{\tilde{\chi}^0_I}}{m_b}.
\eea
One could ask whether the results obtained in section 2 for heavy
Dirac fermions can be applied to the case of neutralinos, which are
Majorana particles. The answer to this question is positive: for the
particular set of Feynman diagrams we have considered, there is
technically no difference between Dirac and Majorana fermions on
internal lines. No ``clashing-arrow'' propagators have to be included,
and no extra combinatoric factors occur.

        Application of the above substitutions gives us the following
neutralino contributions to the Wilson coefficients:
\bea \label{neutralino4}
\delta^{\tilde{\chi}^0} C_4^{(1)}(\mu_0) &=&  
\f{1}{g_2^2 K_{ts}^* K_{tb}} \sum_{A=1}^6 \sum_{I=1}^4 \f{M_W^2}{m_{\tilde{\chi}^0_I}^2} 
        \left( Z^{D_L}_I \right)^*_{A2} \left( Z^{D_L}_I \right)_{A3} H_1^{[4]}(w_{\scs AI}) 
\hspace{1.3cm} \mbox{with~~} w_{\scs AI}= \f{m_{\tilde{d}_A}^2}{m_{\tilde{\chi}^0_I}^2},\\[2mm]
\delta^{\tilde{\chi}^0} C_{7,8}(\mu_0) &=&  
\f{1}{g_2^2 K_{ts}^* K_{tb}} \sum_{A=1}^6 \sum_{I=1}^4 \f{M_W^2}{m_{\tilde{\chi}^0_I}^2} 
\times \nonumber\\ && \hspace{-1.8cm} \times \left\{
  \left( Z^{D_L}_I \right)^*_{A2} \left( Z^{D_L}_I \right)_{A3}
                \left[ H_3^{[7,8]}  (w_{\scs AI}) 
+ \f{\al}{4\pi} \left( H_3^{[7,8]'} (w_{\scs AI}) 
                     + H_3^{[7,8]''}(w_{\scs AI}) 
    \ln \left( \f{\mu_0^2}{m_{\tilde{d}_A}^2}\right) \right) \right]
\right. \nonumber\\ && \hspace{-2.5cm}
+ \left. \f{m_{\tilde{\chi}^-_I}}{m_b} \left( Z^{D_L}_I \right)^*_{A2} \left( Z^{D_R}_I \right)_{A3} 
                \left[ H_4^{[7,8]}  (w_{\scs AI}) 
+ \f{\al}{4\pi} \left( H_4^{[7,8]'} (w_{\scs AI}) 
                     + H_4^{[7,8]''}(w_{\scs AI}) 
    \ln \left( \f{\mu_0^2}{m_{\tilde{d}_A}^2}\right) \right) \right] \right\}, \hspace{1cm} 
\label{neutralino78}
\eea
where~~
$H^{[8]  }_3(x) = H^{[8]  }_1(x)$,~~~
$H^{[8]  }_4(x) = H^{[8]  }_2(x) + \f{1}{2}$,~~~
$H^{[8]' }_{3,4}(x) = H^{[8]' }_{1,2}(x)$,~~~
$H^{[8]''}_{3,4}(x) = H^{[8]''}_{1,2}(x)$,\\
$H^{[7]}_3(x) = -\f{1}{3} H^{[8]}_1(x)$,~~~
$H^{[7]}_4(x) = -\f{1}{3} \left( H^{[8]}_2(x) + \f{1}{2} \right)$~~~
and
\bea
\begin{array}{lclr}
H_3^{[7]'}(x)  &=&    \f{16x^2+28x}{9(1-x)^4} Li_2\left(1-\f{1}{x}\right) 
              \;+\;   \f{-108x^2-358x-38}{81(1-x)^5} \ln x 
              \;+\;   \f{23x^3-765x^2-693x-77}{243(1-x)^4},
&\hspace{1.5cm} 
 (\arabic{equation}) \addtocounter{equation}{1}\\[4mm]
H_3^{[7]''}(x) &=&    \f{4x^2-28x}{9(1-x)^5} \ln x
              \;+\;   \f{-8x^3+42x^2-84x-22}{27(1-x)^4},
&(\arabic{equation}) \addtocounter{equation}{1}\\[4mm]
H_4^{[7]'}(x)  &=&    \f{16x^2+48x}{9(1-x)^3} Li_2\left(1-\f{1}{x}\right)
              \;+\;   \f{-8x^2-68x-8}{9(1-x)^4} \ln x
              \;+\;   \f{-26x^2-54x-4}{9(1-x)^3},
&(\arabic{equation}) \addtocounter{equation}{1}\\[4mm]
H_4^{[7]''}(x) &=&    \f{8x^2-44x}{9(1-x)^4} \ln x
              \;+\;   \f{10x^2-30x-16}{9(1-x)^3}.
&(\arabic{equation}) \addtocounter{equation}{1} \end{array} \nonumber
\eea
Contributions to the primed coefficients can be obtained by
interchanging $Z^{D_L}$ and $Z^{D_R}$ in eqs.~(\ref{neutralino4}) and
(\ref{neutralino78}).\\

        Finally, we turn to contributions from gluino--squark loops,
i.e. from one-loop diagrams with gluinos and two-loop diagrams with
both gluinos and gluons. The heavy fermion and scalar considered in
section 2 are now in the adjoint and fundamental representations of
$SU(3)_{colour}$, respectively. Comments concerning Majorana fermions
we have made in the context of neutralinos apply here as well. The
necessary substitutions in the present case are the following:
$m_{\psi} \to m_{\tilde{g}}$,~ 
$m_{\phi} \to m_{\tilde{d}_A}$,~
$Q_{\psi} \to 0$,~ 
$Q_{\phi} \to -1/3$,~
$\kappa_{\psi} \to 3$,~
$\kappa_{\phi} \to 4/3$~ and
\bea
N \left( S_L^* B_L + \f{m_s}{m_b} S_R^* B_R \right) &=& 
\f{8g_3^2}{3g_2^2 K_{ts}^* K_{tb}} \; \f{M_W^2}{m_{\tilde{d}_A}^2} \;
\left( \Gamma^{D_L} \right)^*_{A2} \left( \Gamma^{D_L} \right)_{A3}
\;\;+\;\; {\cal O}\left(\f{m_s^2}{M_W^2}\right),
\nonumber\\
N S_L^* B_R \f{m_{\psi}}{m_b} &=& 
-\f{8g_3^2}{3g_2^2 K_{ts}^* K_{tb}} \; \f{M_W^2}{m_{\tilde{d}_A}^2} \;
\left( \Gamma^{D_L} \right)^*_{A2} \left( \Gamma^{D_R} \right)_{A3}
\f{m_{\tilde{g}}}{m_b}.
\eea
The resulting contributions to the Wilson coefficients read
\bea \label{gluino4}
\delta^{\tilde{g}} C_4^{(1)}(\mu_0) &=&  
\f{8g_3^2}{3g_2^2 K_{ts}^* K_{tb}} \f{M_W^2}{m_{\tilde{g}}^2} \sum_{A=1}^6  
        \left( \Gamma^{D_L} \right)^*_{A2} \left( \Gamma^{D_L} \right)_{A3} H_5^{[4]}(v_{\scs A}) 
\hspace{1.3cm} \mbox{with~~} v_{\scs A}= \f{m_{\tilde{d}_A}^2}{m_{\tilde{g}}^2},\\[2mm]
\delta^{\tilde{g}} C_{7,8}(\mu_0) &=&  
\f{8g_3^2}{3g_2^2 K_{ts}^* K_{tb}} \f{M_W^2}{m_{\tilde{g}}^2} \sum_{A=1}^6  
\times \nonumber\\ && \hspace{-1.8cm} \times \left\{
  \left( \Gamma^{D_L} \right)^*_{A2} \left( \Gamma^{D_L} \right)_{A3}
                \left[ H_5^{[7,8]}  (v_{\scs A}) 
+ \f{\al}{4\pi} \left( H_5^{[7,8]'} (v_{\scs A}) 
                     + H_5^{[7,8]''}(v_{\scs A}) 
    \ln \left( \f{\mu_0^2}{m_{\tilde{d}_A}^2}\right) \right) \right]
\right. \nonumber\\ && \hspace{-2.2cm} \left.
- \f{m_{\tilde{g}}}{m_b} \left( \Gamma^{D_L} \right)^*_{A2} \left( \Gamma^{D_R} \right)_{A3} 
                \left[ H_6^{[7,8]}  (v_{\scs A}) 
+ \f{\al}{4\pi} \left( H_6^{[7,8]'} (v_{\scs A}) 
                     + H_6^{[7,8]''}(v_{\scs A}) 
    \ln \left( \f{\mu_0^2}{m_{\tilde{d}_A}^2}\right) \right) \right] \right\}, \hspace{1cm} 
\label{gluino78}
\eea
where~~
$H^{[7]}_5(x) = -\f{1}{3} H^{[8]}_1(x)$,~~~
$H^{[7]}_6(x) = -\f{1}{3} \left( H^{[8]}_2(x) + \f{1}{2} \right)$~~~
and
\bea
\begin{array}{lclr}
H_5^{[4]}(x)    &=&    \f{18x^3-27x^2+1}{24(1-x)^4} \ln x
               \;+\;   \f{73x^2-134x+37}{72(1-x)^3}
,& (\arabic{equation}) \addtocounter{equation}{1}\\[4mm]
H_5^{[7]'}(x)  &=&    \f{17x^2+86x-15}{18(1-x)^4} Li_2 \left( 1 - \f{1}{x} \right)
              \;+\;   \f{6x^3+45x^2+66x-5}{12(1-x)^5} \ln^2 x\\[4mm]
       &&     \;+\;   \f{-36x^4-315x^3+1161x^2+751x+23}{162(1-x)^5} \ln x
              \;+\;   \f{-799x^3+1719x^2+10431x-1847}{972(1-x)^4}                       
,& \hspace{1cm}
   (\arabic{equation}) \addtocounter{equation}{1}\\[4mm]
H_5^{[7]''}(x) &=&    \f{18x^3+107x^2+43x}{18(1-x)^5} \ln x
              \;+\;   \f{-5x^3+384x^2+609x+20}{108(1-x)^4}
,&(\arabic{equation}) \addtocounter{equation}{1}\\[4mm]
H_6^{[7]'}(x)  &=&    \f{19x^2+60x-15}{9(1-x)^3} Li_2 \left( 1 - \f{1}{x} \right)
              \;+\;   \f{6x^3+36x^2+48x-5}{6(1-x)^4} \ln^2 x
\\[4mm]&&     \;+\;   \f{-27x^3+106x^2+52x+1}{9(1-x)^4} \ln x
              \;+\;   \f{14x^2+333x-83}{18(1-x)^3}
,&(\arabic{equation}) \addtocounter{equation}{1}\\[4mm]
H_6^{[7]''}(x) &=&    \f{18x^3+80x^2+28x}{9(1-x)^4} \ln x
              \;+\;   \f{55x^2+69x+2}{9(1-x)^3}
,&(\arabic{equation}) \addtocounter{equation}{1}\\[4mm]
H_5^{[8]}(x)   &=&    \f{9x^2-x}{16(1-x)^4} \ln x
              \;+\;   \f{19x^2+40x-11}{96(1-x)^3}
,&(\arabic{equation}) \addtocounter{equation}{1}\\[4mm]
H_5^{[8]'}(x)  &=&    \f{45x^3-1208x^2+901x-570}{96(1-x)^4} Li_2 \left( 1 - \f{1}{x} \right)
              \;+\;   \f{-237x^3-846x^2+282x-95}{32(1-x)^5} \ln^2 x
\\[4mm]&&     \;+\;   \f{2520x^4-10755x^3-10638x^2-6427x-44}{864(1-x)^5} \ln x
              \;+\;   \f{5359x^3-241425x^2+143253x-59251}{5184(1-x)^4}
,&(\arabic{equation}) \addtocounter{equation}{1}\\[4mm]
H_5^{[8]''}(x) &=&    \f{-747x^3-640x^2+43x}{48(1-x)^5} \ln x
              \;+\;   \f{-779x^3-7203x^2-93x+11}{288(1-x)^4}
,&(\arabic{equation}) \addtocounter{equation}{1}\\[4mm]
H_6^{[8]}(x)   &=&    \f{9x^2-x}{8(1-x)^3} \ln x
              \;+\;   \f{13x-5}{8(1-x)^2}
,&(\arabic{equation}) \addtocounter{equation}{1}\\[4mm]
H_6^{[8]'}(x)  &=&    \f{-359x^2+339x-204}{24(1-x)^3} Li_2 \left( 1 - \f{1}{x} \right)
              \;+\;   \f{-78x^3-333x^2+105x-34}{8(1-x)^4} \ln^2 x
\\[4mm]&&     \;+\;   \f{-207x^3-1777x^2+23x-151}{48(1-x)^4} \ln x
              \;+\;   \f{-1667x^2+990x-379}{24(1-x)^3}
,&(\arabic{equation}) \addtocounter{equation}{1}\\[4mm]
H_6^{[8]''}(x) &=&    \f{-126x^3-133x^2+7x}{6(1-x)^4} \ln x
              \;+\;   \f{-553x^2+84x-35}{12(1-x)^3}
.&(\arabic{equation}) \addtocounter{equation}{1} 
\end{array} \nonumber
\eea
Contributions to the primed coefficients can be obtained by
interchanging $\Gamma^{D_L}$ and $\Gamma^{D_R}$ in
eqs.~(\ref{gluino4}) and (\ref{gluino78}).

Our leading-order SSM results agree with
refs.~\cite{BBMR91,CDGG98,CMW96}.  The next-to-leading results are
new, except for the chargino ones, which will be compared with those
of ref.~\cite{CDGG98} in the next section.

\ \\
{\bf 6. Numerical size of the NLO corrections}

In the present section, we shall give a numerical example of the NLO
matching effect in the $B \to X_s \gamma$ branching ratio. From among
many possible extensions of the SM to which our results can be
applied, we choose the Supersymmetric Standard Model, because of its
current popularity. Our present results allow us to make the NLO
prediction for $BR[B \to X_s \gamma]$ in the SSM only in certain
regions of its parameter space. We cannot consider situations where
the gluino mass is close in size to $M_W$ or $m_t$, because two-loop
matching diagrams with no gluons (but only gluinos) have not been
calculated so far.\footnote{
It would require including one-loop gluino corrections
to the Wilson coefficients of four-quark operators, too.}
In addition, we have to assume that $\tan\beta$ is not much
larger than unity. Otherwise, two-loop diagrams containing no QCD
interactions at all would become numerically relevant, and the
analysis would be much more involved.

For simplicity, we shall consider a scenario in which the gluino is
much heavier than {\em all} the other SSM particles, and we shall neglect
all the $1/m_{\tilde{g}}$ effects. This scenario is somewhat different
from the one considered in ref.~\cite{CDGG98}, where some of the other
superpartners were considered heavy as well. 

So long as only the gluino is heavy and decouples, QCD corrections to
loops with other supersymmetric particles are given by two-loop
diagrams with gluons (calculated in the previous sections) and by
two-loop diagrams with strong quartic squark couplings (found in
appendix C). One-loop gluino corrections to the Wilson coefficients of
four-quark operators vanish in this limit. However, we need to take
into account the fact that the theory with decoupled gluino is no
longer supersymmetric. Consequently, the usual tree-level
expressions for chargino and neutralino couplings are affected by
${\cal O}(\al)$ corrections, which do {\em not} vanish in the
$m_{\tilde{g}} \to \infty$ limit.

When the gluino mass $m_{\tilde{g}}$ is much larger than masses of all
the other SSM particles, one should, in principle, perform the
decoupling in two steps. First, one should match the complete SSM with
the effective theory built out of all the SSM particles except for the
gluino. This matching should be performed at the renormalization scale
$\mu_{\tilde{g}}$ that should be of the same order as
$m_{\tilde{g}}$. Next, one should perform the RGE evolution of all the
couplings in the obtained effective theory down to the scale $\mu_0$
that should be of the same order as $M_W$ or $m_t$. At this scale,
all the remaining SSM particles with masses ${\cal O}(M_W,m_t)$ are
decoupled, and the Wilson coefficients $C_4$, $C_7$ and $C_8$ are
found.

The scales $\mu_{\tilde{g}}$ and $\mu_0$ can be set equal to each
other so long as $\ln \f{ \mu_{\tilde{g}}}{ \mu_W}$ is not much larger
than unity. This can happen even in situations where the gluino is
still heavy enough, for instance when $m_{\tilde{g}} \sim 700$~GeV,
and all the remaining SSM particles have masses below, say, 350~GeV.
The results can then be written in a compact form, because no RGE
evolution occurs between $\mu_{\tilde{g}}$ and $\mu_0$. We shall take
advantage of this opportunity in our example here.

Once the gluino is decoupled, chargino and neutralino couplings
$X^{U_{L,R}}$, $Z^{D_{L,R}}$ are related to the other parameters of
the model as follows (cf. eq.~(2.12) of ref.~\cite{CMW96}):
\bea
X_I^{U_L} &=& g_2 \left[ - a_g V_{I1}^* \Gamma^{U_L} 
+ a_Y V_{I2}^* \Gamma^{U_R} \f{M_U}{\sqrt{2} M_W \sin \beta} \right] K, \\[2mm]
\label{xur}
X_I^{U_R} &=& g_2 a_Y U_{I2} \Gamma^{U_L} K \f{M_D}{\sqrt{2} M_W \cos \beta}, \\[2mm]
Z_I^{D_L} &=& -\f{g_2}{\sqrt{2}} \left[ 
a_g \left(-N_{I2}^* + \f{1}{3}\tan\theta N_{I1}^* \right) \Gamma^{D_L} 
+ a_Y N_{I3}^* \Gamma^{D_R} \f{M_D}{M_W \cos \beta} \right], \\[2mm]
Z_I^{D_R} &=& -\f{g_2}{\sqrt{2}} \left[ 
a_g \f{2}{3}\tan\theta N_{I1} \Gamma^{D_R} 
+ a_Y N_{I3} \Gamma^{D_L} \f{M_D}{M_W \cos \beta} \right],
\eea
where
\be \label{aYag}
a_g = 1 + \f{\al(\mu_0)}{\pi} \left( \ln \f{m_{\tilde{g}}}{\mu_0} - \f{7}{12} \right)
\mbox{~~~~~and~~~~~} 
a_Y = 1 - \f{\al(\mu_0)}{\pi} \left( \ln \f{m_{\tilde{g}}}{\mu_0} - \f{1}{4} \right).
\ee
In the above equations, all the parameters (including the matrices
$X^{U_{L,R}}$ and $Z^{D_{L,R}}$) are assumed to be
$\overline{MS}$-renormalized in dimensional {\em regularization}, in
the effective theory containing no gluino. Setting $a_g$ and $a_Y$ to
unity, we would obtain relations for parameters of the full SSM,
$\overline{MS}$-renormalized in dimensional {\em reduction}. Explicit
values of $a_g$ and $a_Y$ have been obtained by performing simple
one-loop matching between these two theories.

One might wonder whether the appearance of $a_g$ and $a_Y$ is the only
effect of gluino decoupling. Potentially, interactions different from
those of the SSM could be generated at one loop. A more detailed
examination of the relevant one-loop diagrams leads to the conclusion
that other effects indeed occur, but they are never relevant to the
NLO QCD corrections to $BR[B \to X_s \gamma]$.

Once $a_Y$ and $a_g$ from eq.~(\ref{aYag}) are included, we are able
to verify that our results for the NLO corrections to chargino loops
(eqs.~(\ref{chargino78})--(\ref{chargino.end}), (\ref{echquartic}) and
(\ref{gchquartic})) agree with those of ref.~\cite{CDGG98}, so long as
the gluino is assumed to be much heavier than all the other SSM
particles. When performing the comparison, one needs to take into
account that our results are expressed in terms of
$\overline{MS}$-renormalized masses, while the on-shell squark masses
were used in ref.~\cite{CDGG98}.

The last worry one could have in the context of the SSM is whether
two-loop effects involving no couplings proportional to $\al$, but
only to the large Yukawa coupling $Y_t$, could be of a numerical
importance similar to the NLO QCD corrections. This can potentially
happen even in the small $\tan\beta$ regime. No definite answer to
this question can be given until the SM calculation of ref.~\cite{S98}
is generalized to the SSM case.\footnote{
  The smallness of the net effect in the SM is partly due to accidental
  cancellations, which may not take place in the SSM.  } 
However, the main purpose of the present section is only to demonstrate
that ${\cal O}(\al)$ effects can be sizable by themselves, independently
of the magnitude of two-loop higgson/higgsino contributions.\\

Having made all the necessary assumptions, we are now ready to collect
the results from section 5 and appendix C to test the size of ${\cal
  O}(\al)$ corrections to $BR[B \to X_s \gamma]$ for some particular
values of the SSM parameters. A relatively simple set of SSM
parameters, which is allowed by all the experimental constraints
(including $B \to X_s \gamma$), can be chosen as follows:
\bea
\tan \beta &=& 3, \hspace{4.5cm}
m_{\tilde{\scs W}}/\mu = -2, \nonumber \\ 
m_{h^{\pm}} &=& 100~{\rm GeV}, \hspace{3cm}
m_{gluino} = 700~{\rm GeV}, \nonumber \\ 
(M_{\tilde{\chi}^{\pm}})_{11} &=& 140~{\rm GeV} 
\mbox{~~~~ (the lighter chargino mass)}, \nonumber \\
(M^2_{\tilde{u}})_{AB} &=& (\delta_{AB} - \delta_{A6}\delta_{B6}) \times (350~{\rm GeV})^2
                         + \delta_{A6}\delta_{B6} \times (110~{\rm GeV})^2, \nonumber \\
\Gamma^U_{AB} &\simeq& \delta_{AB} 
            + (\cos 25^o - 1) (\delta_{A3}\delta_{B3} + \delta_{A6}\delta_{B6}) 
             + \sin 25^o (\delta_{A6}\delta_{B3} - \delta_{A3}\delta_{B6}), \nonumber \\
\Gamma^D &\simeq& \mbox{~~(flavour-diagonal matrix)}. \nonumber 
\eea
The first three parameters determine the chargino masses and mixing
angles. Down-squark masses, neutralino masses and mixing angles are
irrelevant here, because we have assumed approximate vanishing of
flavour violation in the down squark mass matrix. Consequently, all the
neutralino contributions to the Wilson coefficients $C_{4,7,8}$
vanish.  As far as the relevant SM parameters are concerned, we take
the same values as in ref.~\cite{CMM97}, i.e. $m_{t,pole} = 175$~GeV 
\linebreak

\begin{tabular}{|rcl|rcl|rcl|}
\hline
\rule[-2mm]{0mm}{8mm}
$\delta^{SM} C_7^{(0)}(M_W)$ &=& --0.197 &
$\delta^{SM} C_8^{(0)}(M_W)$ &=& --0.098 & && \\
\rule[-2mm]{0mm}{8mm}
$\delta^{H^-} C_7^{(0)}(M_W)$ &=& --0.279 & 
$\delta^{H^-} C_8^{(0)}(M_W)$ &=& --0.211 & && \\
\rule[-2mm]{0mm}{8mm}
$\delta^{\tilde{\chi}^-} C_7^{(0)}(M_W)  $ &=&  ~0.272 &
$\delta^{\tilde{\chi}^-} C_8^{(0)}(M_W)  $ &=&  ~0.148 & && \\
\hline
\rule[-2mm]{0mm}{8mm}
$\delta^{SM} C_7^{(1)}(M_W)$ &=& --2.49 &
$\delta^{SM} C_8^{(1)}(M_W)$ &=& --2.22 &
$\delta^{SM} C_4^{(1)}(M_W)$ &=& --0.42 \\
\rule[-2mm]{0mm}{8mm}
$\delta^{H^-} C_7^{(1)}(M_W)$ &=& ~5.24 &
$\delta^{H^-} C_8^{(1)}(M_W)$ &=& ~2.85 &
$\delta^{H^-} C_4^{(1)}(M_W)$ &=& ~0.02 \\
\rule[-2mm]{0mm}{8mm}
$\delta^{\tilde{\chi}^-} C_7^{(1)}(M_W)$ &=& --0.24 &
$\delta^{\tilde{\chi}^-} C_8^{(1)}(M_W)$ &=&  ~0.18 &
$\delta^{\tilde{\chi}^-} C_4^{(1)}(M_W)$ &=&  ~0.08 \\
\rule[-2mm]{0mm}{8mm}
$\delta^{\tilde{\chi}^-}_q C_7^{(1)}(M_W)$ &=& --2.22 &
$\delta^{\tilde{\chi}^-}_q C_8^{(1)}(M_W)$ &=& --1.49 & && \\
\hline
\end{tabular}
\begin{center}
Table 1.~~ Numerical contributions to the Wilson coefficients 
           in the considered example
\end{center}

\noindent
and $\al(M_Z) = 0.118$. For the Wolfenstein parameters, we choose
$\lambda = 0.22$,~ $A=0.83$,~ $\rho=0$ and $\eta=0.3$. Such $\rho$ and
$\eta$ are roughly in the middle of the allowed region in the SM. They
are acceptable also in the SSM, for the parameters we have specified
above.

The contributions to the Wilson coefficients obtained at $\mu_0 = M_W$
are given in table 1. The ${\cal O}(\al \ln m_{\tilde g}^2/\mu_0^2)$
terms in chargino and neutralino couplings have been included both in
the leading and next-to-leading contributions to the Wilson
coefficients, as should be done in the SSM with decoupled gluino.

Having found the SSM matching conditions numerically, we evaluate
$BR[B \to X_s \gamma]$ according to the formulae of ref.~\cite{CMM97}
and using the same input parameters. However, we choose a different
photon energy cut-off $\delta = 0.90$ \cite{KN98}, change
$\alpha_{em}(m_b)$ to $\alpha_{em}^{on\;\;shell} \simeq \f{1}{137}$ in
the overall normalization \cite{CM98}, and include the $1/m_c$
corrections \cite{one.ov.mc}.  The final result for $BR[B \to X_s
\gamma]$ is
\be
BR[B \to X_s \gamma] = 3.15 \times 10^{-4},
\ee
which is equal to the central value of the CLEO measurement quoted in
the introduction. On the other hand, if we did not include the
two-loop gluonic SUSY contributions (i.e. if we set $\delta^{H^-}
C_{7,8}^{(1)}(M_W)$ and $\delta^{\tilde{\chi}^-} C_{7,8}^{(1)}(M_W)$
to zero), we would obtain $BR[B \to X_s \gamma] = 3.69 \times
10^{-4}$.  The difference between the two results is close in size to
the present experimental uncertainty. Such a size of the two-loop SUSY
effect is rather generic for light supersymmetric particles, for which
the leading SUSY contributions to $B \to X_s \gamma$ are of the same
order as the SM one. Thus, QCD corrections to superpartner loops are
expected to be important when making a meaningful comparison with
experiment in such cases.

Obviously, satisfying the experimental $B \to X_s \gamma$ bound for
light superpartners requires a certain adjustment of the SSM
parameters.  In our case, the adjusted parameter was the lighter
chargino mass.  The experimental $B \to X_s \gamma$ bound would be
satisfied within $1\sigma$ for this mass ranging from 110 to 175~GeV,
i.e. no real fine-tuning was necessary.

For larger $\tan \beta$, SUSY contributions to $B \to X_s \gamma$ can
be much larger than the SM one. Then, the NLO SUSY effects are much
more important than in the example presented above. However,
satisfying the experimental $B \to X_s \gamma$ constraint requires real
fine-tuning then.

Our results for the two-loop SUSY contributions to $B \to X_s \gamma$
could be used in scans over the SSM parameter space. The allowed
regions in this space would change their position for light
superpartners.  However, our feeling is that making such scans at
present would be somewhat premature. The NLO SUSY contributions to $B
\to X_s \gamma$ can become qualitatively important only when either
light superpartners are discovered or their existence is almost
excluded by data combined from many experiments.  One of these options
will certainly be realized when the LHC starts collecting data. At
that time, one might appreciate the usefulness of the long analytic
formulae contained in the present paper.

\ \\
{\bf 7. Summary}

We have calculated matching conditions for operators mediating the $b
\to s \gamma$ and \linebreak $b \to s\;gluon$ transitions in a large
class of extensions of the Standard Model. Both the leading one-loop
diagrams and the gluonic corrections to them have been included.
Taking the Supersymmetric Standard Model as an example, we have
checked that QCD corrections to new physics contributions can be close
in size to the present experimental uncertainty, even when the SSM
parameters are not really fine-tuned. A similar situation is expected to
occur in other theories of new physics containing exotic particles at
the electroweak scale.

The main purpose of our paper was to present complete analytic
formulae for the NLO Wilson coefficients in a possibly generic
extension of the SM. From these results, we could reproduce the known
matching conditions in the Standard Model, the Two-Higgs Doublet
Model, and for the chargino contributions in the SSM. New results were
obtained for the left--right-symmetric model, as well as for
neutralino and gluino contributions in the SSM. Our SSM results form a
major contribution to the complete SSM calculation. They become
complete by themselves in the heavy-gluino limit and for $\tan \beta$
of order unity.

\ \\
{\bf Acknowledgements}

C.B and J.U. thank F.~Krauss, K.~Schubert and G.~Soff for helpful
discussions.  M.M. thanks Andrzej Buras and Paolo Gambino for useful
advice. This work has been supported in part by the German
Bundesministerium f{\"u}r Bildung und Forschung under contracts
06~DD~823 (J.U.) and 06~TM~874 (M.M). M.M. has been supported
in part by the DFG project Li~519/2-2, as well as by the Polish
Committee for Scientific Research under grant 2~P03B~014~14,
1998-2000.

\newpage \noindent 
{\bf Appendix A}

        In this appendix, we summarize several useful identities
involving colour generators and the Clebsch--Gordan coefficients
$C_{ijk}$. Thanks to them, all the colour factors in section 2 could
have been expressed only in terms of $\kappa_{\psi}$, $\kappa_{\phi}$,
and $\xi$ (cf. eqs.~(\ref{kappa}) and (\ref{xi})):
\bea
T^a_{(\phi)ij} T^a_{(\psi)kl} C_{jlm} &=& 
       \f{1}{2}\left(\f{4}{3}-\kappa_{\phi}-\kappa_{\psi}\right) C_{ikm},
\\
T^a_{(\phi)ij} C_{jkl} T^a_{lm} &=& 
       \f{1}{2}\left(\f{4}{3}+\kappa_{\phi}-\kappa_{\psi}\right) C_{ikm},
\\
T^a_{(\psi)kj} C_{ijl} T^a_{lm} &=& 
       \f{1}{2}\left(\f{4}{3}+\kappa_{\psi}-\kappa_{\phi}\right) C_{ikm},
\\
C^*_{ijn} T^a_{(\phi)il} C_{ljm} &=& \f{3}{8} \xi
     \left( \f{4}{3} + \kappa_{\phi} - \kappa_{\psi} \right) T^a_{nm},
\\
C^*_{jin} T^a_{(\psi)il} C_{jlm} &=& \f{3}{8} \xi
     \left( \f{4}{3} + \kappa_{\psi} - \kappa_{\phi} \right) T^a_{nm},
\\
C^*_{ijn} \left( T^a_{(\phi)} T^b_{(\phi)} \right)_{il} C_{ljm} &=& 
\f{\xi}{8} \left[ \kappa_{\phi} \delta_{ab} \delta_{nm}
+ \f{3i}{2} \left( \f{4}{3} + \kappa_{\phi} - \kappa_{\psi} \right) f_{abc} T^c_{nm}
+ \f{3}{5} \eta_{\phi} d_{abc} T^c_{nm} \right],
\\
C^*_{jin} \left( T^a_{(\psi)} T^b_{(\psi)} \right)_{il} C_{jlm} &=& 
\f{\xi}{8} \left[ \kappa_{\psi} \delta_{ab} \delta_{nm}
+ \f{3i}{2} \left( \f{4}{3} + \kappa_{\psi} - \kappa_{\phi} \right) f_{abc} T^c_{nm}
+ \f{3}{5} \eta_{\psi} d_{abc} T^c_{nm} \right], \hspace{4mm} 
\eea
where
\bea
\eta_{\phi} &=& 3 (\kappa_{\phi} - \kappa_{\psi})^2  
            + \f{3}{2} \kappa_{\phi} - \f{7}{2} \kappa_{\psi} -\f{2}{3},
\\
\eta_{\psi} &=& 3 (\kappa_{\psi} - \kappa_{\phi})^2  
            + \f{3}{2} \kappa_{\psi} - \f{7}{2} \kappa_{\phi} -\f{2}{3}.
\eea
All these identities can be derived from the basic constraint given in
eq.~(\ref{constraints}).

        For completeness, let us quote the standard identities for the
fundamental representation, too:
\be 
T^a T^b = \f{i}{2} f_{abc} T^c + \f{1}{2} d_{abc} T^c 
   + \f{1}{6} \delta_{ab} \mbox{\large 1},
\ee
\be
f_{abc} f_{abd} = 3 \delta_{cd},
\hspace{2cm}
d_{abc} d_{abd} = \f{5}{3} \delta_{cd},
\hspace{2cm}
d_{aab} = 0.
\ee

\ \\
{\bf Appendix B}

Here, we present explicit formulae for the functions $f_i(x)$,
$g_i(x)$, $h_i(x)$, $j_i(x)$ and $k_i(x)$ introduced in sections 2 and
3. All these expressions are available in Mathematica format via
anonymous ftp from
ftp://feynman.t30.physik.tu-muenchen.de/pub/preprints/tum-hep-321-98.functions.m~.
They read:
\input functions.dat

\ \\
{\bf Appendix C}

This appendix is devoted to presenting the SSM matching contributions
originating from the quartic squark vertex proportional to the strong
coupling constant $\al$:
\bea
\delta^{\tilde{\chi}^-}_q C_7^{(1)}(\mu_0) &=&  \f{1}{g_2^2 K_{ts}^* K_{tb}} 
\;\; \sum_{A,B,C=1}^6 \;\;\ \sum_{I=1}^2 \;\; \f{M_W^2}{m_{\tilde{\chi}^-_I}^4} \; 
P^{\scs U}_{\scs AB} m^2_{\tilde{u}_B} P^{\scs U}_{\scs BC} \; 
\left( \ln \f{ m^2_{\tilde{u}_B}}{\mu_0^2} - 1 \right)
\times \nonumber\\ && \times 
\left\{ \left( X^{U_L}_I \right)^*_{A2} \left( X^{U_L}_I \right)_{C3} 
\left[ -q_1(z_{\scs AI},z_{\scs CI})+\f{2}{3} q_2(z_{\scs AI},z_{\scs CI}) \right] 
\right. \nonumber\\ && \hspace{-0.7cm} + \left. \f{m_{\tilde{\chi}^-_I}}{m_b} 
\left( X^{U_L}_I \right)^*_{A2} \left( X^{U_R}_I \right)_{C3} 
\left[ -q_3(z_{\scs AI},z_{\scs CI})+\f{2}{3} q_4(z_{\scs AI},z_{\scs CI}) \right] 
\right\}, \label{echquartic} \\[4mm] 
\delta^{\tilde{\chi}^-}_q C_8^{(1)}(\mu_0) &=&  \f{1}{g_2^2 K_{ts}^* K_{tb}} 
\;\; \sum_{A,B,C=1}^6 \;\;\ \sum_{I=1}^2 \;\; \f{M_W^2}{m_{\tilde{\chi}^-_I}^4} \; 
P^{\scs U}_{\scs AB} m^2_{\tilde{u}_B} P^{\scs U}_{\scs BC} \; 
\left( \ln \f{ m^2_{\tilde{u}_B}}{\mu_0^2} - 1 \right)
\times \nonumber\\ && \hspace{-2.5cm} \times 
\left\{ \left( X^{U_L}_I \right)^*_{A2} \left( X^{U_L}_I \right)_{C3} 
q_2(z_{\scs AI},z_{\scs CI}) + \f{m_{\tilde{\chi}^-_I}}{m_b} 
\left( X^{U_L}_I \right)^*_{A2} \left( X^{U_R}_I \right)_{C3} 
q_4(z_{\scs AI},z_{\scs CI}) 
\right\}, \label{gchquartic} \\[4mm] 
-3 \delta^{\tilde{\chi}^0}_q C_7^{(1)}(\mu_0) &=&  
\delta^{\tilde{\chi}^0}_q C_8^{(1)}(\mu_0) \nonumber\\ &=& \f{1}{g_2^2 K_{ts}^* K_{tb}} 
\;\; \sum_{A,B,C=1}^6 \;\;\ \sum_{I=1}^4 \;\; \f{M_W^2}{m_{\tilde{\chi}^0_I}^4} \; 
P^{\scs D}_{\scs AB} m^2_{\tilde{d}_B} P^{\scs D}_{\scs BC} \; 
\left( \ln \f{ m^2_{\tilde{d}_B}}{\mu_0^2} - 1 \right)
\times \nonumber\\ && \hspace{-2.5cm} \times 
\left\{ \left( Z^{D_L}_I \right)^*_{A2} \left( Z^{D_L}_I \right)_{C3} 
q_2(w_{\scs AI},w_{\scs CI})  + \f{m_{\tilde{\chi}^0_I}}{m_b} 
\left( Z^{D_L}_I \right)^*_{A2} \left( Z^{D_R}_I \right)_{C3} 
q_4(w_{\scs AI},w_{\scs CI}) \right\}, \hspace{1cm}\\[4mm]
\delta^{\tilde{g}}_q C_7^{(1)}(\mu_0) &=&  -\f{8 g_3^2}{9 g_2^2 K_{ts}^* K_{tb}} 
\; \f{M_W^2}{m_{\tilde{g}}^4}  \;\; \sum_{A,B,C=1}^6 \;\;
P^{\scs D}_{\scs AB} m^2_{\tilde{d}_B} P^{\scs D}_{\scs BC} \; 
\left( \ln \f{ m^2_{\tilde{d}_B}}{\mu_0^2} - 1 \right)
\times \nonumber\\ && \hspace{-2.5cm} \times 
\left\{ \left( \Gamma^{D_L} \right)^*_{A2} \left( \Gamma^{D_L} \right)_{C3} 
q_2(v_{\scs A},v_{\scs C})  - \f{m_{\tilde{g}}}{m_b} 
\left( \Gamma^{D_L} \right)^*_{A2} \left( \Gamma^{D_R} \right)_{C3} 
q_4(v_{\scs A},v_{\scs C}) \right\},\\[4mm]
\delta^{\tilde{g}}_q C_8^{(1)}(\mu_0) &=&  \f{3 g_3^2}{g_2^2 K_{ts}^* K_{tb}} 
\; \f{M_W^2}{m_{\tilde{g}}^4}  \;\; \sum_{A,B,C=1}^6 \;\;
P^{\scs D}_{\scs AB} m^2_{\tilde{d}_B} P^{\scs D}_{\scs BC} \; 
\left( \ln \f{ m^2_{\tilde{d}_B}}{\mu_0^2} - 1 \right)
\times \nonumber\\ && \times 
\left\{ \left( \Gamma^{D_L} \right)^*_{A2} \left( \Gamma^{D_L} \right)_{C3} 
\left[ q_1(v_{\scs A},v_{\scs C}) - \f{1}{9} q_2(v_{\scs A},v_{\scs C}) \right] 
\right. \nonumber\\ && \hspace{-0.5cm} - \left. \f{m_{\tilde{g}}}{m_b} 
\left( \Gamma^{D_L} \right)^*_{A2} \left( \Gamma^{D_R} \right)_{C3} 
\left[ q_3(v_{\scs A},v_{\scs C}) - \f{1}{9} q_4(v_{\scs A},v_{\scs C}) \right] 
\right\},
\eea
where 
$P^{\scs U} = \Gamma^{U_L} \Gamma^{U_L\dagger}-\Gamma^{U_R} \Gamma^{U_R\dagger}$
~and~
$P^{\scs D} = \Gamma^{D_L} \Gamma^{D_L\dagger}-\Gamma^{D_R} \Gamma^{D_R\dagger}$. 
The mass ratios are denoted as before: 
 $z_{\scs AI}= m_{\tilde{u}_A}^2/m_{\tilde{\chi}^-_I}^2$,
~$w_{\scs AI}= m_{\tilde{d}_A}^2/m_{\tilde{\chi}^0_I}^2$ ~and
~$v_{\scs A}= m_{\tilde{d}_A}^2/m_{\tilde{g}}^2$.

The explicit expressions for the functions $q_i(x,y)$ read
\bea
\begin{array}{lclr}
q_1(x,y) &=& \f{2}{3(x-y)} \left[ \f{x^2 \ln x}{(1-x)^4} - \f{y^2 \ln y}{(1-y)^4} \right]
+ \f{2 x^2 y^2 + 5 x^2 y + 5 x y^2 - x^2 - y^2 - 22 x y + 5 x + 5 y + 2}{9(1-x)^3(1-y)^3}, 
& \hspace{1.7cm} (\arabic{equation}) \addtocounter{equation}{1}\\[4mm]
q_2(x,y) &=& \f{2}{3(x-y)} \left[ \f{x \ln x}{(1-x)^4} - \f{y \ln y}{(1-y)^4} \right]
+ \f{-x^2 y^2 + 5 x^2 y + 5 x y^2 +2 x^2 + 2 y^2 - 10 x y - 7 x - 7 y + 11}{9(1-x)^3(1-y)^3}, 
&(\arabic{equation}) \addtocounter{equation}{1}\\[4mm]
q_3(x,y) &=& \f{4}{3(x-y)} \left[ \f{x^2 \ln x}{(1-x)^3} - \f{y^2 \ln y}{(1-y)^3} \right]
+\f{-6xy+2x+2y+2}{3(1-x)^2(1-y)^2}, 
&(\arabic{equation}) \addtocounter{equation}{1}\\[4mm]
q_4(x,y) &=& \f{4}{3(x-y)} \left[ \f{x \ln x}{(1-x)^3} - \f{y \ln y}{(1-y)^3} \right]
+\f{-2xy-2x-2y+6}{3(1-x)^2(1-y)^2}. 
&(\arabic{equation}) \addtocounter{equation}{1}\\[4mm]
\end{array}
\nonumber
\eea

\setlength {\baselineskip}{0.2in}
 
\end{document}